\documentclass[structabstract]{aa}  
\usepackage{graphicx}
\usepackage{txfonts}
\pdfoutput=1

\usepackage{natbib}
\bibpunct{(}{)}{;}{a}{}{,}

\newcommand{\cmd}{\mbox{$\mbox{cm}^{-2}$}}
\newcommand{\kms}{\mbox{$\mbox{km} \, \mbox{s}^{-1}$}}
\newcommand{\lsun}{\mbox{$\rm L_\odot$}}
\newcommand{\msun}{\mbox{${\rm M}_\sun$}}

\newcommand{\Nhtwo}{$N_{\rm H_2}$}

\newcommand{\vlsr}{\mbox{$v_{\rm LSR}$}}
\newcommand{\tmb}{\mbox{$T_{\mbox{\tiny MB}}$}}

\begin{document}
%\doublespacing
%
   \title{The formation of the W43 complex: constraining its atomic-to-molecular transition and searching for colliding clouds}

\author{F. Motte\inst{1}
          \and Q.~Nguy$\tilde{\hat{\rm e}}$n~Lu{\hskip-0.65mm\small'{}\hskip-0.5mm}o{\hskip-0.65mm\small'{}\hskip-0.5mm}ng\inst{2}
          \and N. Schneider\inst{3}
          \and F. Heitsch\inst{4}
          \and S. Glover\inst{5}
          \and P. Carlhoff\inst{6}
          \and{T. Hill}\inst{7,1}          
          \and S. Bontemps\inst{3}
          \and P. Schilke\inst{6}
          \and{F. Louvet}\inst{1}
          \and{M. Hennemann}\inst{8}
          \and{P. Didelon}\inst{1}
           \and H. Beuther \inst{8} 
         }

  \institute{Laboratoire AIM Paris-Saclay, CEA/IRFU - CNRS/INSU - Universit\'e Paris Diderot, Service d'Astrophysique, B\^at. 709, CEA-Saclay, F-91191 Gif-sur-Yvette Cedex, France, 
              \email{frederique.motte@cea.fr}
      \date{Received 2013; accepted 2013}
           \and Canadian Institute for Theoretical Astrophysics, University of Toronto, 60 St. George Street, Toronto, ON M5S~3H8, Canada
           \and OASU/LAB-UMR~5804, CNRS/INSU - Universit\'e Bordeaux 1, 2 rue de l'Observatoire, BP 89, F-33270 Floirac, France 
           \and Department of Physics and Astronomy, University of North Carolina Chapel Hill, Phillips Hall, Chapel Hill, NC 27599-3255, USA
           \and Universit\"at Heidelberg, Zentrum f\"ur Astronomie, Institut f\"ur Theoretische Astrophysik, Albert-Ueberle-Str. 2, 69120 Heidelberg, Germany
           \and Physikalisches Institut, Universit\"at zu K\"oln, Z\"ulpicher Str. 77, D-50937 K\"oln, Germany
           \and Joint ALMA Observatory, Alonso de Cordova 3107, Vitacura, Santiago, Chile.
           \and Max-Planck-Institut f\"ur Astronomie, K\"onigsstuhl 17, 69117 Heidelberg, Germany
        }

\abstract
  % context heading (optional)
  % {} leave it empty if necessary  
   {Numerical simulations have explored the possibility to form molecular clouds through either a quasi-static, self-gravitating mechanism or the collision of gas streams or lower-density clouds. They also quantitatively predict the distribution of matter at the transition from atomic to molecular gases.}
  % aims heading (mandatory)
   {We aim to observationally test these models by studying the environment of W43, a molecular cloud complex recently identified near the tip of the Galactic long bar.}
  % methods heading (mandatory)
   {Using Galaxy-wide \ion{H}{I} and $^{12}$CO 1--0 surveys we searched for gas flowing toward the W43 molecular cloud complex. We also estimated the \ion{H}{I} and H$_{2}$ mass surface densities to constrain the transition from atomic to molecular gas around and within W43.}
  % results heading (mandatory)
   {We found three cloud ensembles within the position-velocity diagrams of $^{12}$CO and \ion{H}{I} gases. They are separated by $\sim$20~$\kms$ along the line of sight and extend into the $^{13}$CO velocity structure of W43. Since their velocity gradients are consistent with free-fall, they could be nearby clouds attracted by, and streaming toward, the W43 $\sim$10$^7~\msun$ potential well. We show that the \ion{H}{I} surface density, $\Sigma_{\ion{H}{I}} = 45-85~\msun\,$pc$^{-2}$, does not reach any threshold level but increases when entering the 130\,pc-wide molecular complex previously defined. This suggests that an equilibrium between H$_2$ formation and photodissociation has not yet been reached. The H$_2$-to-\ion{H}{I} ratio measured over the W43 region and its surroundings, $R_{{\rm H_{2}}}\sim3.5\pm_{2}^{3}$, is high, indicating that most of the gas is already in molecular form in W43 and in structures several hundreds of parsecs downstream along the Scutum-Centaurus arm.}
  % conclusions heading (optional), leave it empty if necessary 
{The W43 molecular cloud complex may have formed, and in fact may still be accreting mass from the agglomeration of clouds. Already in the molecular-dominated regime, most of these clouds are streaming from the Scutum-Centaurus arm. This is in clear disagreement with quasi-static and steady-state models of molecular cloud formation.} 

   \keywords{ISM: atoms, ISM: clouds, ISM: evolution, ISM: individual (W43), ISM: structure, stars: formation}

\titlerunning{Atomic-to-molecular transition and putative colliding clouds in W43}
\authorrunning{Motte, Nguyen Luong et al.}
\maketitle
%________________________________________________________________

%%%%%%%%%%%%%%%%%%%%%%% 1. Introduction %%%%%%%%%%%%%%%%%%
\section{Introduction}

%%%%%%%%%%%%%%%%%%%% 1.1 Converging flows %%%%%%%%%%%%%%%%%%
The term `converging flows', although applicable to all mechanisms of cloud formation, generally refers to the convergence of \ion{H}{I} streams that can naturally be driven by local instabilities in the disk such as those due to gravity, supernova explosions, or spiral shocks \citep[e.g.][]{KoIn00,heitsch05,vazquez07}. For the past decade, numerical models have investigated the capability of such colliding flows (Warm Neutral Medium, $\sim$$6 000$~K) to form cold structures (Cold Neutral Medium, $\sim$70~K) through shocks  \citep[e.g.][]{vazquez96,HePe99,ballesteros99}. However, it is only recently that 3D models have simulated the thermal transition from the atomic to the molecular phase (\ion{H}{I} $\rightarrow$ H$_2$) with realistic heating and cooling functions \citep[e.g.][]{AuHe05,GlML07}. Several groups aim at studying in detail the formation of molecular clouds with sufficient resolution to model star formation. Recent 3D numerical simulations now include gravity, magnetic fields, thermal and dynamical instabilities \citep[e.g.][]{banerjee09, hartmann12} and in some cases treat the chemical evolution of the gas \citep{HeHa08, clark12, InIn12}.

Massive giant molecular clouds, or cloud complexes, may alternatively form from the collision/agglomeration of clouds \citep[e.g.][]{BlSh80,dobbs08}. Clouds are higher density and more structured media than diffuse \ion{H}{I} streams, and so they are mainly composed of cold \ion{H}{I} and H$_2$ gas. Cloud-cloud collisions are expected to be the dominant phenomenon in high-density regions of galaxies \citep[e.g.][]{dobbs08} as well as during the late-stage evolution of molecular clouds formed through \ion{H}{I} converging flows  \citep[e.g.][]{vazquez10}. Several groups are studying the formation, evolution, and disruption of massive clouds by spiral waves, bar potentials, and satellite galaxies with hydrodynamic simulations of galaxies able to resolve molecular clouds \citep[e.g.][]{bournaud10, DP13}.

Both colliding flows and cloud-cloud collisions naturally explain the observed rapid onset of star formation once the cloud has formed, which has proven problematic in the past \citep[$<$$3-10 \times 10^6$~yr,][]{ballesteros99,hartmann01,roman09}. In these scenarios, molecular clouds are never in a kinematic equilibrium state, as part of the cloud collapses while part of it disperses. A few models have described the formation of molecular clouds by assuming equilibrium between the formation of H$_2$ molecules and their photodissociation \citep[e.g.][]{AnWa93,krumholz09}. In short, these authors applied a simplified version of the chemical modeling of photodissociation regions \citep[e.g.][]{vDiBl86} to a spherical geometry. However, such an equilibrium may never be reached since its timescale should be $1-3\times 10^7$~yr or even longer \citep{MLG12}. Non steady-state models have been developed using turbulence and/or colliding streams to enhance and accelerate the formation of H$_2$ molecules behind shocks \citep[e.g.][]{heitsch08,hennebelle08,KoIn00,clark12}. These two families of models make different predictions, for the conversion of \ion{H}{I} into H$_2$, especially in terms of the lifetime of the process and the mixing of the H$_2$ and \ion{H}{I} gases. Indeed, steady-state models assume that an equilibrium between the \ion{H}{I} and H$_2$ formation/destruction is rapidly reached. The \ion{H}{I} and H$_2$ gases are also assumed to be mutually exclusive, with a sharp transition from \ion{H}{I} to H$_2$-dominated media when entering the cloud \citep[e.g][]{krumholz09}. As a consequence, equilibrium models predict the existence of a threshold for the atomic gas surface density, at $\Sigma_{\ion{H}{I}} \simeq 10~\msun\,$pc$^{-2}$ for solar metallicity gas according to \cite{krumholz09}. In contrast, non-steady state models start with an accumulation of \ion{H}{I} gas that translates into a mass surface density above this classical threshold. The H$_2$ formation rate, high at the beginning of the molecular cloud formation process, continuously slows down as the cloud becomes fully molecular \citep[e.g.][]{clark12}. The \ion{H}{I} surface density thus decreases over time, first sharply, then slower and slower, eventually reaching this equilibrium state.

%%%%%%%%%%%%%%%%%%%%%%% 1.3 W43 Paper I %%%%%%%%%%%%%%%%%%
W43 should be a perfect testbed to investigate the formation of molecular clouds and star formation in the framework of dynamical scenarios. At only 5.5~kpc from us, W43 is among the most extreme molecular cloud complexes of the Milky Way \citep{nguyen11b,zhang14}. It is massive, $M_{\mbox{\tiny total}}\sim 6\times 10^6~\msun$\footnote{
	The mass and diameter values given in \cite{nguyen11b} have been recalculated to account for the refinement of W43 distance, from 6~kpc to 5.5~kpc, by \cite{zhang14}.} 
within an equivalent diameter of $\sim$130~pc, and is highly concentrated into dense star-forming sites \citep{nguyen11b}.  Despite a velocity dispersion FWHM of $\sim$$22.3~\kms$, the W43 molecular cloud complex is a coherent and gravitationally bound ensemble of clouds. Moreover, W43 has the potential to form starburst clusters in the near future \citep[SFR$\, \sim 0.1~\msun\, \mbox{yr}^{-1}$,][]{nguyen11b,louvet14}. Its densest parts correspond to the Galactic mini-starburst cloud W43-Main \citep{motte03} recently mapped by the \emph{Herschel}/Hi-GAL and HOBYS key programs \citep{molinari10,motte10}. \citet{nguyen13} identified within W43-Main two dense filamentary clouds categorized as ridges following the definition of \citet{hill11} and \citet{hennemann12}. In short, ridges are elongated clouds with very high column density, \Nhtwo$\,>10^{23}~\cmd$, over several squared parsecs \citep[see also][]{nguyen11a,motte12}. \citet{nguyen13} proposed and \cite{louvet14} showed that the W43-MM1 and W43-MM2  ridges are progenitors of young massive clusters. W43 is located near the meeting point of the Scutum-Centaurus (or Scutum-Crux) Galactic arm and the Galactic long bar, a dynamically complex region where collisions are expected \citep{nguyen11b, carlhoff13}. Several massive stellar associations, called red supergiant clusters, have also been identified in the $1\degr-6\degr$ neighborhood of W43 \citep[see][and references therein]{gonz12}.

In this paper, we attempt to trace back the formation of molecular clouds in the W43 complex. Employing the database described in Sect.~\ref{s:obs}, we look for gas streams and cloud ensembles at the outskirts of the atomic and molecular complex and characterize its atomic envelope (see Sect.~\ref{s:results}). Section~\ref{s:discu} follows the transition from atomic to molecular media and discusses the collision probability of gas streams and clouds. Finally, Sect.~\ref{s:conc} concludes that the equilibrium relations of the  \ion{H}{I} into H$_2$ transition do not apply in W43 and that this extreme cloud complex may have been created by the collision of molecular clouds.

%%%%%%%%%%%%%%%%%%%%% 2. Observations %%%%%%%%%%%%%%%%%%%
\section{Observations}
\label{s:obs}

The atomic gas data for W43 are taken from the Very Large Array (VLA) Galactic Plane Survey (VGPS\footnote{
	The full spectral cubes of the VGPS database are available as fits files at \emph{http://www.ras.ucalgary.ca/VGPS/}. The VLA and GBT are facilities of the National Radio Astronomy Observatory.},
\citealt{stil06}). VGPS is a survey of the 21 cm continuum and line emission from neutral atomic hydrogen, \ion{H}{I}, performed throughout the Galactic plane ($18\degr < l < 67\degr$, $|b| < 1.3\degr$) by the VLA and combined with short-spacing information obtained with the Green Bank Telescope (GBT). This data set has a spatial resolution of $1\arcmin\times1\arcmin$ and a velocity resolution of $1.56~\kms$ spanning the range -120 to $170~\kms$. 

To trace the low-density molecular gas, we used the $^{12}$CO 1--0 data from the Galactic plane survey made with the CfA 1.2~m telescope\footnote{
	The CfA telescope is  the 1.2~mm Millimeter-Wave Telescope at the Center for Astrophysics, Harvard University.}  
\citep{dame01}. These data have a  spatial resolution of $450\arcsec$ and a velocity resolution of $0.65~\kms$ spanning the velocity range -0.5 to $271~\kms$. Note that $^{13}$CO data of e.g. the GRS survey \citep{jackson06} trace higher density gas than $^{12}$CO and are thus less relevant for our study of the transition from atomic to molecular gas. The main characteristics of the data sets used here are listed in Table~\ref{table:obs}.

Low-density cloud structures surrounding the W43 complex as defined by \cite{nguyen11b} are expected to display optically thin \ion{H}{I} and $^{12}$CO lines. In contrast, these two lines should be self-absorbed toward the densest parts of W43. \ion{H}{I} lines will also be absorbed by the continuum free-free emission of \ion{H}{II} regions, seen as pointlike sources within the densest parts of W43.

%%%%%%%%%%%%%%%%%%%%% Table 1 %%%%%%%%%%%%%%%%%%%%%%%
\begin{table}[htbp]
\caption{Main observational parameters.}
\label{table:obs} 
\centering 
\begin{tabular}{cccccc}
\hline
\hline
Survey 						&  Frequency	&    HPBW  & $\Delta {\rm v}_{\mbox{\tiny res}}$ 	& 1$\sigma$~rms\\   
$\diagup$Tracer   				&         [GHz] 	&    [\arcsec]    & [$\kms$] 						& [K$\,\kms$]   \\   
\hline
VGPS$\diagup$\ion{H}{I}	& 1.420		&   60		& 1.56						& 1.80   \\
CfA$\diagup$$^{12}$CO 1--0    	& 115.271 	& 450		& 0.65 						& 0.22   \\
\hline
\end{tabular}
\end{table}

%%%%%%%%%%%%%%%%%%%%%%%%%%%%%%%%%%%%%%%%%%%%%%%%%%%%%%%%%
%%%%%%%%%%%%%%%%%%%%% 3. Results and Analysis %%%%%%%%%%%%%%%%%%%%%%%%
\section{Results and analysis}
\label{s:results}

Here we discuss the velocity range covered by the $^{12}$CO and \ion{H}{I} envelope of W43 (see Sect.~\ref{s:coline}), characterize the \ion{H}{I} envelope of W43 (see Sect.~\ref{s:hienv}), and search for signatures of colliding flows or clouds (see Sect.~\ref{s:coflow}).

%%%%%%%%%%%%%%%%%%%%% 3.1 Velocity range %%%%%%%%%%%%%%%%%%%%%%%%
\subsection{Velocity range of the $^{12}$CO/\ion{H}{I} envelope of W43}
\label{s:coline}
 
\begin{figure}[tp]
\vskip -0.2cm
\centering
\includegraphics[width=9cm]{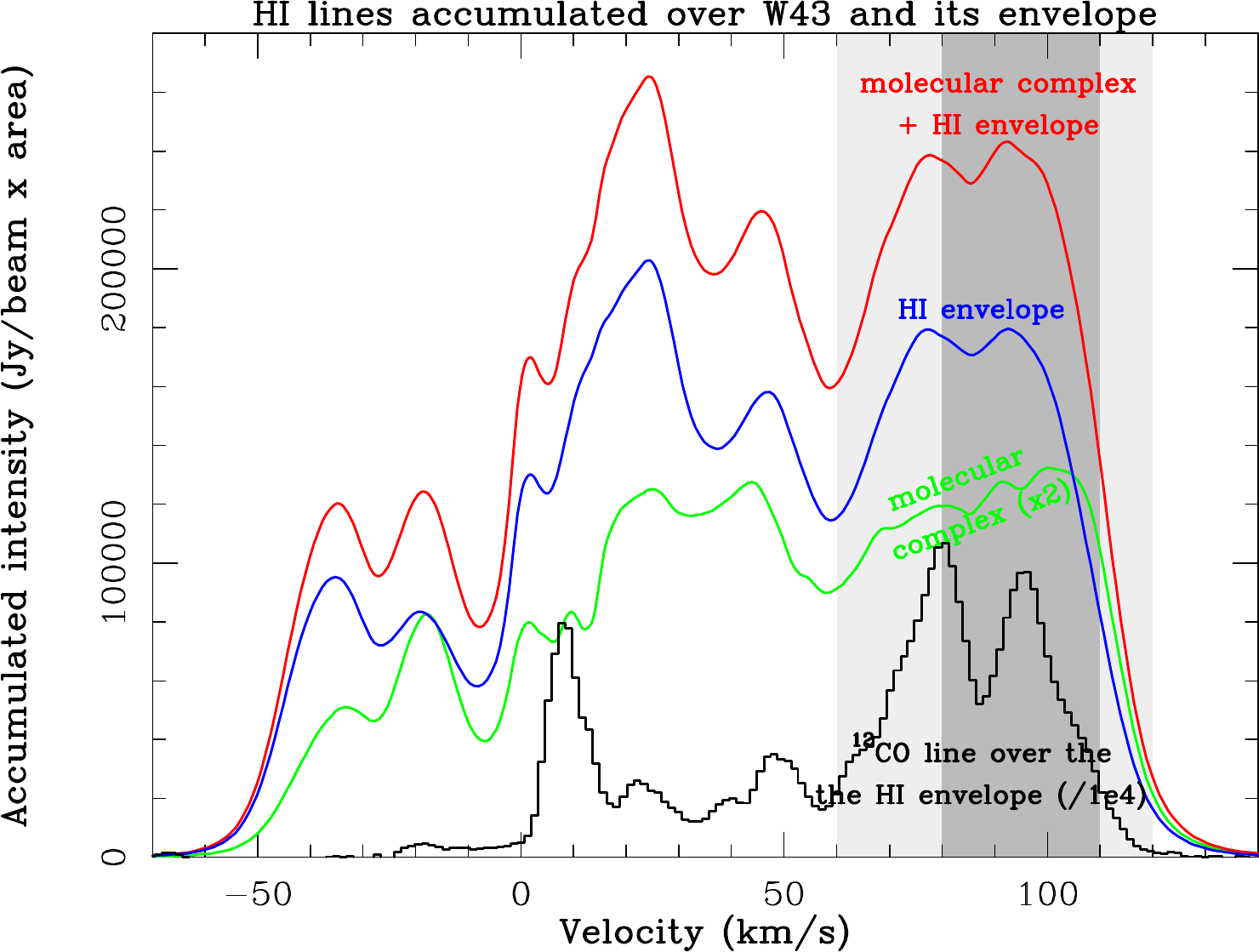} 
\caption{ 
W43 \ion{H}{I} lines (curves) well delimited by the $60-120~\kms$ velocity range and displaying a double-peak shape similar to the $^{12}$CO 1--0 lines (black histogram). \ion{H}{I} spectra result from the sum over all observed beams in the $1.8\degr\times0.8\degr$ area covering the W43 molecular complex (green curve, $\times 3$, within the white rectangle of Fig.~\ref{fig:co_hi_intmap}), those within its \ion{H}{I} envelope (blue curve, in between the white continuous and dashed rectangles of Fig.~\ref{fig:co_hi_intmap}) and the $3\degr\times 2\degr$ area covering both components (red curve, within the white dashed rectangle of Fig.~\ref{fig:co_hi_intmap}). The main and complete velocity ranges of W43 defined by \citet{nguyen11b} are indicated by the light  and dark gray shaded regions, respectively.}
\label{fig:co_hi_spec}
\vskip -0.5cm
\end{figure}

\begin{figure*}[htbp]
$\begin{array}{l}
\includegraphics[width=18cm,angle=0]{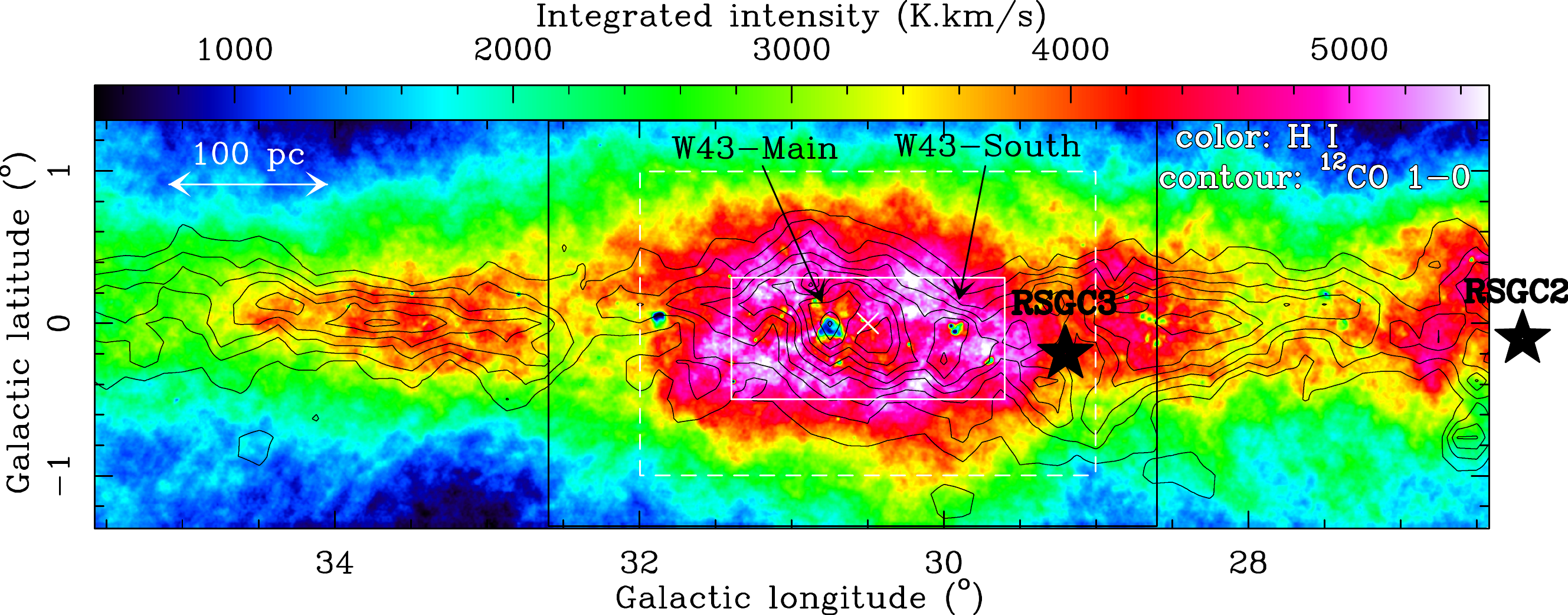} 
\end{array}$
\caption{The Galactic plane surroundings of the W43 molecular complex as seen in \ion{H}{I} (color) and $^{12}$CO~1--0 (contours) integrated over $60-120~\kms$, the complete velocity range of W43. The contours go from 40 to 200~K$\,\kms$ with a step of 20~K$\,\kms$, and always are monotonously increasing, never displaying $^{12}$CO dips (see Fig.~\ref{fig:hih2_surfdens}b). The white boxes outline the extent of the W43 molecular complex (continuous lines) and its \ion{H}{I} envelope (dashed lines) as defined in \citet{nguyen11b}. The black box locates the largest ring used in the  $R_{{\rm H_{2}}}$ diagram (see Fig.~\ref{fig:gasratio1}) and the white cross its center. Stars pinpoint the red supergiant clusters/associations (RSGC2 and RSGC3) identified in the neighborhood.}
\label{fig:co_hi_intmap}
\end{figure*}

Figure~\ref{fig:co_hi_spec} displays the shape of the \ion{H}{I} lines of Table~\ref{table:obs} summed over areas covering the W43 molecular complex, its $^{12}$CO/\ion{H}{I} envelope, and the combination of the two.  According to \citet{nguyen11b}, the W43 molecular cloud complex extends in Galactic longitude from $l=29.6\degr$ to $31.4\degr$ and in Galactic latitude from $b=-0.5\degr$ to $b=0.3\degr$. If the $^{12}$CO/\ion{H}{I} envelope of the complex is also included, the combined region extends from $l=29\degr$ to $32\degr$ and $|b|<1\degr$. The corresponding areas are outlined in Fig.~\ref{fig:co_hi_intmap} as white and white-dashed rectangles, and lines integrated over these areas are displayed in Fig.~\ref{fig:co_hi_spec} as green and red curves, respectively. Comparison of the \ion{H}{I} and $^{12}$CO spectra shown in Fig.~\ref{fig:co_hi_spec} with the $^{13}$CO spectra discussed by \cite{carlhoff13} situates the $0-60~\kms$ velocity components in the foreground of W43 (mostly within the Sagittarius arm) and the -$70-0~\kms$ components in its background. A complete disentangling of CO velocity components in the foreground and background of W43 can be found in \citet[][see especially their Fig.~5]{carlhoff13}.

%----------------------------------------------- Figure 1: HI+12CO Spectra-----------------------------------------------------------
%------------------------------------------------------------------------------------------------------------------------------------------------------

In Fig.~\ref{fig:co_hi_spec}, the $^{12}$CO and \ion{H}{I} lines of Table~\ref{table:obs}, observed toward W43 (green curve) and its surrounding envelope (blue curve) are broad and structured. They happen to be rather well delimited by the 60 and $120~\kms$ velocities, marking out the the so-called `complete velocity range' of W43. We recall that two velocity ranges have been defined for W43 using $^{13}$CO 1--0 data from the GRS survey \citep{jackson06} and the higher-resolution $^{13}$CO 2--1 image of the W43-HERO\footnote{
	The W43-HERO (W43 Hera/EmiR Observation) project is an IRAM 30~m large program, whose description and data can be accessed at: http://www.iram-institute.org/EN/content-page-292-7-158-240-292-0.html}
 large program \citep{nguyen11b, carlhoff13}. The main velocity range ($\vlsr=80-110~\kms$) traces the central part of the $^{13}$CO cloud while the complete range ($60-120$~$\kms$) also includes peripheral lower-density clouds. Since the \ion{H}{I} and $^{12}$CO 1--0 lines trace lower density gas than $^{13}$CO, the velocity range over which they are detected is generally larger. 
 
\begin{figure*}[hbtp]
  \centering
$\begin{array}{ll}
 \includegraphics[angle=0,width=8.3 cm]{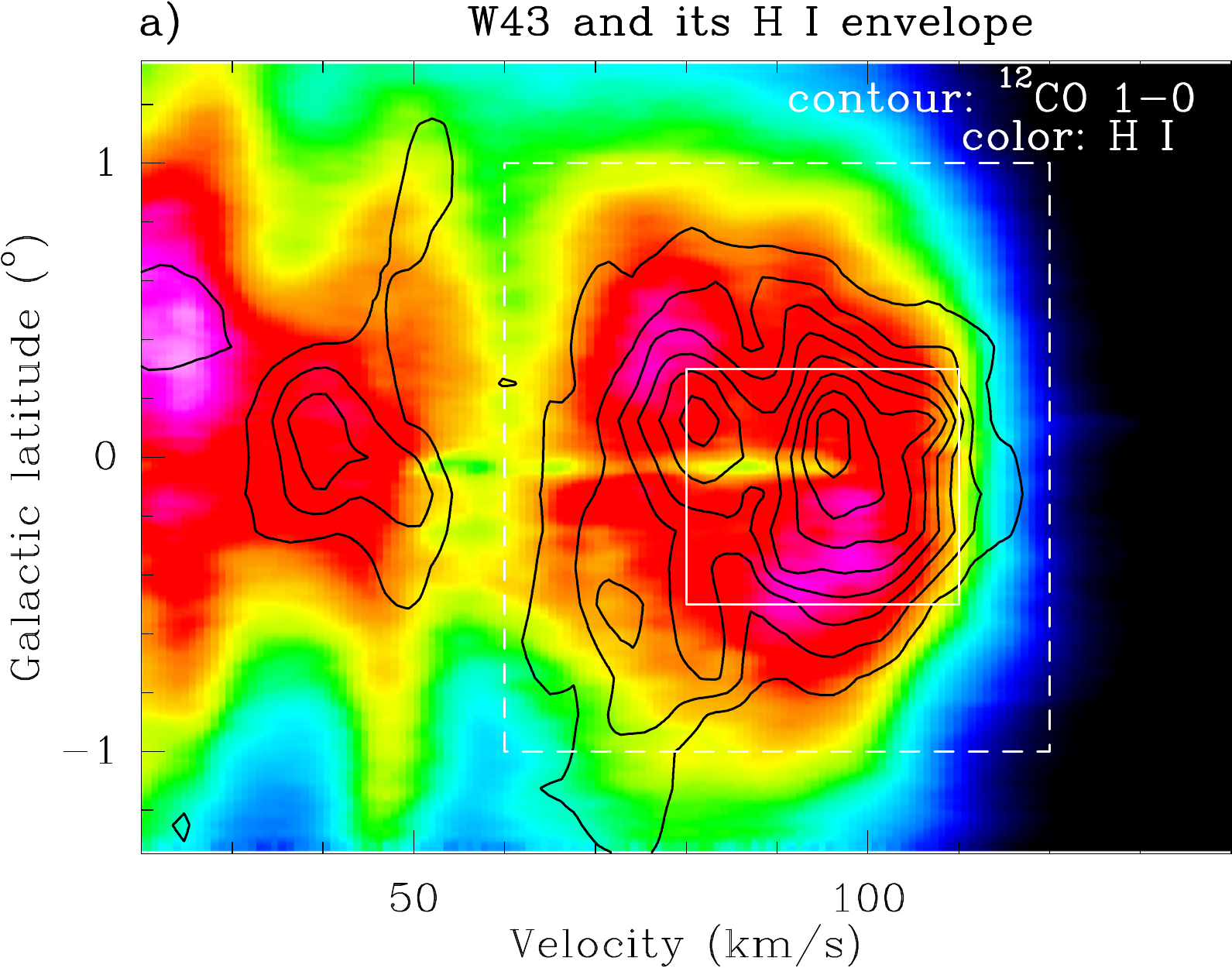}
 \hskip 0.2cm
 \includegraphics[angle=0,width=9.1 cm]{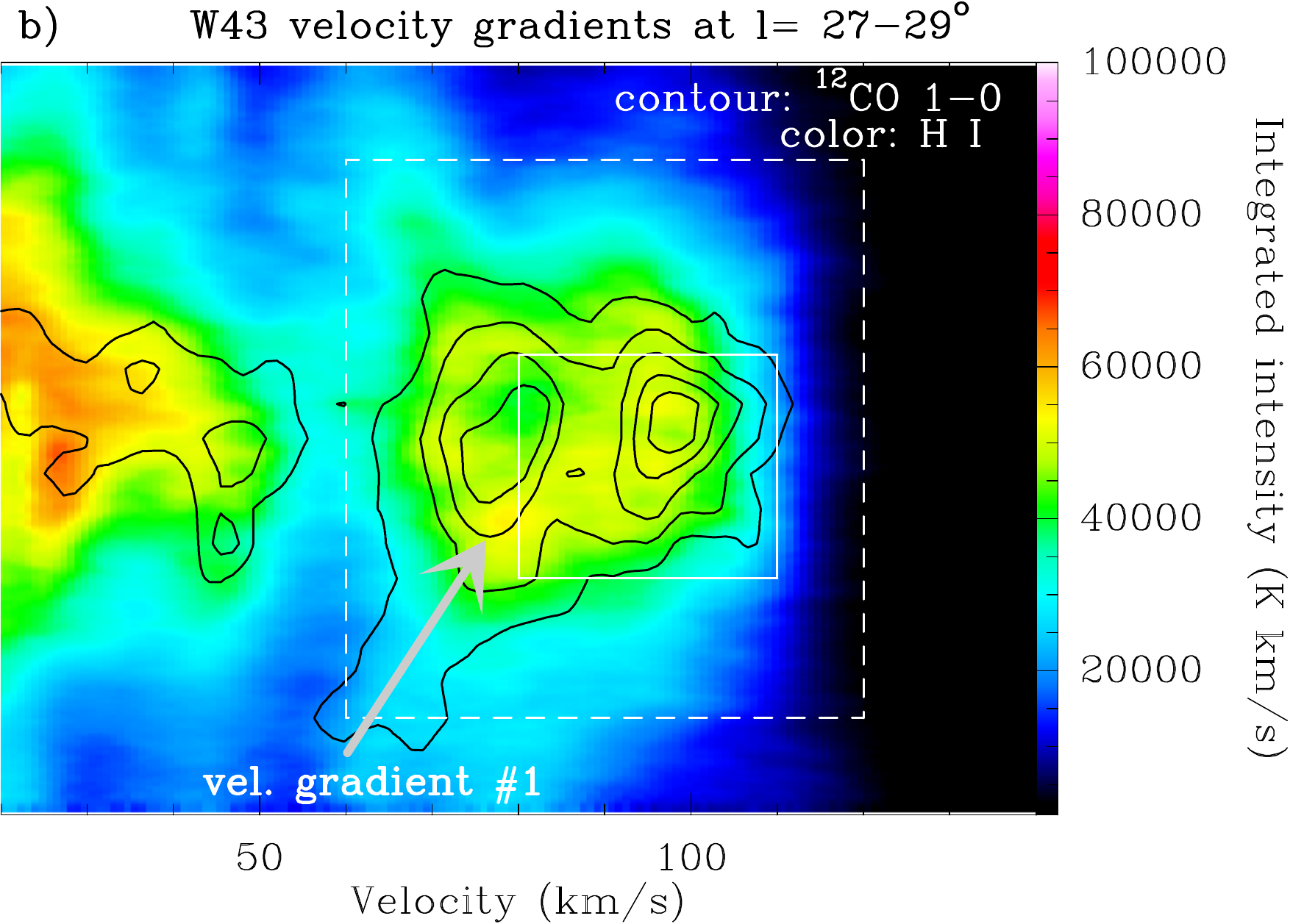}
 \end{array}$
\caption{
Position-velocity diagrams confirming the presence of two velocity components in W43 (see Sect.~\ref{s:coline}) and revealing the \ion{H}{I}/$^{12}$CO~1--0 cloud ensemble~\#1  (white arrow, see Sect.~\ref{s:coflow}). The distributions of \ion{H}{I} (color) and $^{12}$CO~1--0 (contours) are plotted against Galactic latitude, kept as the $y$ axis, as in Fig.~\ref{fig:co_hi_intmap}. The \ion{H}{I} and $^{12}$CO~1--0 lines have been summed {\bf a} over the $l=29\degr-32\degr$ longitude range corresponding to W43 and its  \ion{H}{I} envelope and {\bf b} over the $l=27\degr-29\degr$ longitude range covering part of the \ion{H}{I}/$^{12}$CO cloud ensembles presented in Sect.~\ref{s:coflow}. The contours run from 400 to 2000~K$\,\kms$ with a step of 400~K$\,\kms$. The white boxes represent the latitude extent and velocity range of the W43 molecular complex (continuous lines) and its \ion{H}{I} envelope (dashed lines).  
}
\label{fig:co_hi_pvdiag1}
\end{figure*}

\begin{figure*}[hbtp]
  \centering
$\begin{array}{l}
\includegraphics[angle=0, height=10cm, width=17 cm]{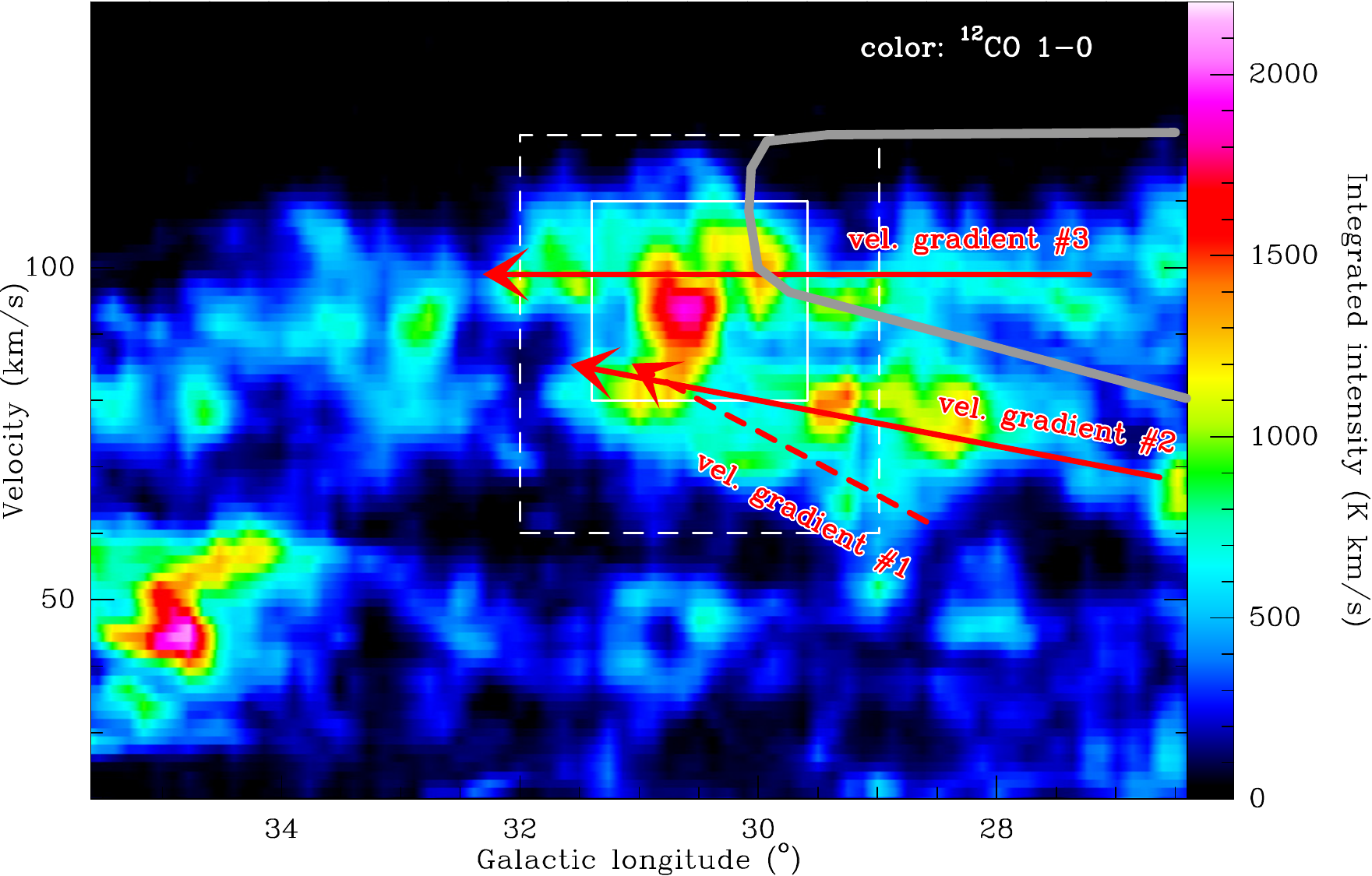}
\end{array}$
\caption{
Position-velocity diagram unfolding the two velocity components of W43 into the $^{12}$CO~1--0 cloud ensembles~\#2 and \#3 (arrows, see Sect.~\ref{s:coflow}). The distributions of $^{12}$CO~1--0 (color) are plotted against the Galactic longitude, kept as the $x$ axis, as in Fig.~\ref{fig:co_hi_intmap}. The $^{12}$CO~1--0 lines have been summed over the $b=-1\degr$ to $b=1\degr$ latitude range corresponding to the W43 complex and its \ion{H}{I} envelope. The white boxes represent the longitude extent and velocity range of the W43 molecular complex (continuous lines) and its \ion{H}{I} envelope (dashed lines). The gray thick curve indicates the theoretical location of the Scutum-Centaurus arm in the model by \citet{vallee08}.
}
\label{fig:co_hi_pvdiag2}
\end{figure*}

\begin{figure*}[hbtp]
  \centering
 \includegraphics[angle=0,height=8.cm, width=15.9 cm]{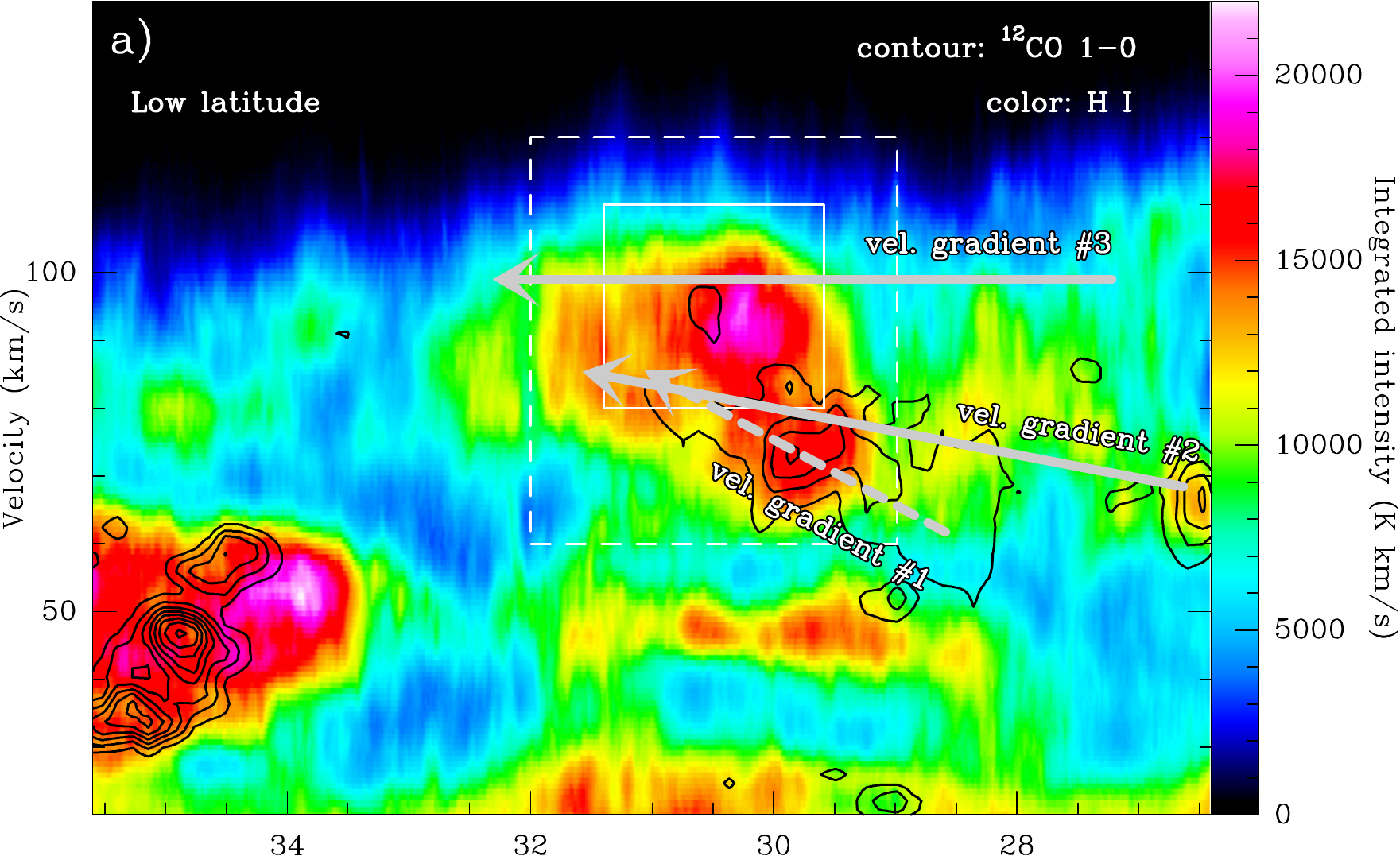} 
 \includegraphics[angle=0,height=8.cm, width=15.9 cm]{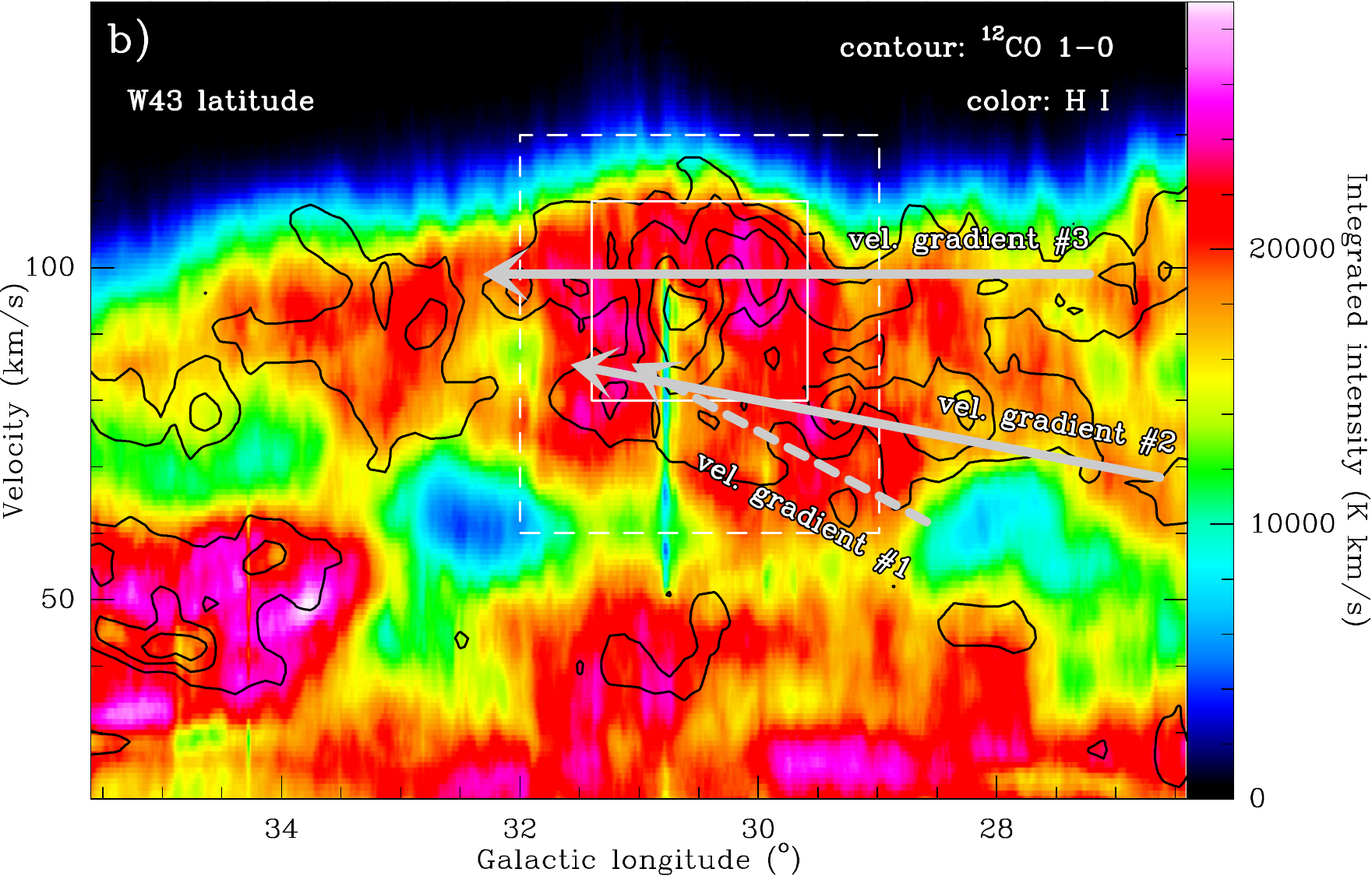}
\caption{
Position-velocity diagrams investigating the $^{12}$CO~1--0/\ion{H}{I} cloud ensembles \#1--\#3 identified in Figs.~\ref{fig:co_hi_pvdiag1}-\ref{fig:co_hi_pvdiag2} (white arrows, see also Sect.~\ref{s:coflow}) at various latitudes. The distributions of the \ion{H}{I} (color) and $^{12}$CO~1--0 (contours) are plotted against the Galactic longitude,  kept as the $x$ axis, as in Fig.~\ref{fig:co_hi_intmap}. The \ion{H}{I} and $^{12}$CO~1--0 lines have been summed in {\bf a} over $b=-1.4\degr$ to $b=-0.7\degr$ to highlight negative-latitude cloud ensembles and in {\bf b} over the $b=-0.5\degr$ to $b=0.3\degr$ latitude range of the W43 molecular complex. The contours of {\bf a} go from 150 to 1200~K$\,\kms$ with a step of 150~K$\,\kms$. Those of {\bf b} run from 300 to 1200~K$\,\kms$, with a step of 300~K$\,\kms$. The white boxes represent the longitude extent and velocity range of the W43 molecular complex (continuous lines) and its \ion{H}{I} envelope (dashed lines). The negative-latitude cloud ensemble \#1 can easily be identified in {\bf a} and the other two cloud ensembles can be seen in {\bf b}. }
\label{fig:co_hi_pvdiag3}
\end{figure*}

The \ion{H}{I} lines of the envelope surrounding W43 (blue curve) sum up in the $60-120~\kms$ range of Fig.~\ref{fig:co_hi_spec} into two main velocity components, leading to a combined {\rm FWHM}$~\sim 60~\kms$. These two velocity peaks are also observed in the $^{12}$CO~1--0 line arising from the envelope of W43 (black histogram) and so do not appear to be due to \ion{H}{I} self-absorption. Indeed, the shape of \ion{H}{I} and CO lines at the highest positive and negative latitudes of the VGPS and CfA data, at $|b|>1\degr$, are as complicated as the spectra displayed in Fig~\ref{fig:co_hi_spec}. The variation of the line shapes when entering the envelope of W43 or W43 itself also do not suggest that they gradually become optically thick. These line shapes provide some evidence that the \ion{H}{I} line is optically thin throughout the envelope of W43 and that at least some of the emission from the W43 complex itself is also optically thin. However, this still needs to be confirmed, and will be quantified using a higher-resolution \ion{H}{I} emission map with the VLA as well as \ion{H}{I} absorption measurements toward radio sources (Bihr et al., in prep.). Assuming that the \ion{H}{I} line shape observed for the \ion{H}{I} envelope also applies to the W43 molecular complex, one can make a crude estimate of the optical thickness of \ion{H}{I} in W43. The lines observed in Fig.~\ref{fig:co_hi_spec} display a similar shape over a wide range of velocities, suggesting that the total \ion{H}{I} flux integrated over the $1.8\degr\times 0.8\degr$ area of W43 is globally underestimated, but by no more than 20\%. Section~\ref{s:hienv} and Fig.~\ref{fig:gasratio1} more precisely indicate the extent of the W43 central region where the \ion{H}{I} line is largely optically thick. Given the above results, we interpret the complicated shape of the \ion{H}{I} and $^{12}$CO lines of Fig.~\ref{fig:co_hi_spec} as reflecting the complex velocity pattern of the W43 region rather than optical thickness. As further discussed in Sect.~\ref{s:coflow} and illustrated in Figs.~\ref{fig:co_hi_pvdiag1}-\ref{fig:co_hi_pvdiag3}, the two velocity peaks observed in Fig.~\ref{fig:co_hi_spec} correspond to $^{12}$CO/\ion{H}{I} cloud ensembles at different velocities.

In the following, we integrate all lines over the $60-120~\kms$ velocity range. With a smaller integration range, the velocity investigations and comparison of the \ion{H}{I} and H$_2$ surface densities made in Sects.~\ref{s:coflow} and \ref{s:discu} are qualitatively unchanged. As for the absolute values of the $^{12}$CO and \ion{H}{I} integrated intensities and thus the H$_2$ and \ion{H}{I} gas masses, they only decrease by $\sim$30\% and $\sim$40\%, respectively, for a $80-110~\kms$ integration range.

%----------------------------------------------- Figure 2: HI+12CO on large scale---------------------------------------------------------
%------------------------------------------------------------------------------------------------------------------------------------------------------

%%%%%%%%%%%%%%%%%%%%% 3.2 HI envelope %%%%%%%%%%%%%%%%%%%%%%%%%%%
\subsection{A clear \ion{H}{I} envelope around W43}
\label{s:hienv}

Figure~\ref{fig:co_hi_intmap} displays the $^{12}$CO 1--0 and \ion{H}{I} images of the W43 molecular complex and its surroundings, integrated over $\vlsr=60-120~\kms$. The observed \ion{H}{I} and $^{12}$CO structures correlate quite well along the Galactic plane, out to the outskirts of the W43 complex, which is outlined by a white rectangle: roughly at $l>31\degr$ and $l<29.75\degr$. Conversely, the \ion{H}{I} and $^{12}$CO are anti-correlated in the densest parts of the molecular complex. This corresponds to areas where the $^{12}$CO~1--0  brightness increases from 140~K$\,\kms$ to 200~K$\,\kms$ in Fig.~\ref{fig:co_hi_intmap} (i.e.\ within the seventh contour) while the \ion{H}{I} emission decreases from $\sim$6000~K$\,\kms$ (pink ring in Fig.~\ref{fig:co_hi_intmap}) down to $\sim$1000~K$\,\kms$ (blue pixels toward W43-Main). Figure~\ref{fig:co_hi_intmap} also confirms that the $^{12}$CO and \ion{H}{I} gases are confined by the midplane pressure of the Milky Way with a width at half maximum of about $0.8\degr$ and $2\degr$, respectively.

Observationally, H$_2$ and \ion{H}{I}  are known to be intimately linked on all scales, from those of large molecular complexes \citep[e.g.][]{BlTh80,elmegreen87} to the scales of individual filamentary clouds \citep[e.g.][]{wannier83,ballesteros99}. Position and velocity coincidences for  \ion{H}{I} and CO have been noted for various clouds and complexes \citep[e.g.][]{BlTh80,elmegreen87}. Halos of warm \ion{H}{I} gas have been found, generally asymmetrically surrounding molecular structures, in the close outskirts of 10~pc filaments/clouds of the nearby Gould Belt clouds \citep[see][and references therein]{wannier83, ballesteros99, lee12}. Similar asymmetrical halos were observed by \citet{williams95} for the high-mass star-forming molecular cloud complex Rosette and its filaments. 

The \ion{H}{I} envelope of the W43 molecular complex itself displays an elliptical shape that symmetrically surrounds the molecular complex (see Fig.~\ref{fig:co_hi_intmap} of the present paper and Fig.~8 of \citealt{nguyen11b}). This envelope, with a median diameter of $\sim$270~pc, is far enough from the internal UV fields of the \ion{H}{II} regions or OB clusters associated with W43 to feel little influence from them; they would otherwise distort its shape. It also lies far enough from the center of the complex to avoid regions where the cold \ion{H}{I} gas co-mixed with molecular gas creates \ion{H}{I} self-absorption features that could distort the \ion{H}{I} envelope emission. Since the \ion{H}{I} envelope has its center close to the Galactic plane, the Galactic interstellar radiation field bathing W43 is expected to be rather isotropic and able to create a symmetrical envelope. The powerful radiation field of RSGC3 could itself distort the W43 envelope at low Galactic longitude if it were not located a few hundred parsecs closer to the Galactic Center  (see Fig.~\ref{fig:co_hi_intmap} and Sect.~\ref{s:historySF}). Note that \ion{H}{I} envelopes of molecular complexes centered on the Galactic plane are not always symmetric (Schneider, priv.\ com.). The large size and simple geometry of the W43 \ion{H}{I} envelope is therefore rare \citep[see, however,][]{Andersson92}, making  W43 an excellent laboratory to track the transition from the atomic to the molecular medium (see Sects.~\ref{s:hisaturation}--\ref{s:hitoh2}). It also suggests that other processes, such as the formation of molecular clouds in the W43 complex, could be invoked to explain its symmetry on large scales. As proposed in Sect.~\ref{s:collidingclouds}, gas could accumulate in front of the long bar and collide to form W43, allowing a symmetric distribution of its \ion{H}{I} and H$_2$ gases on hundred parcsec scales. Note that position offsets between \ion{H}{I} and H$_2$ are present at smaller scales, as mentioned in Sect.~\ref{s:coflow} below.

%----------------------------------------------- Figure 3: HI+12CO PV diag LAT-----------------------------------------------------------
%------------------------------------------------------------------------------------------------------------------------------------------------------

%----------------------------------------------- Figure 4: HI+12CO PV diag LONG---------------------------------------------------------
%------------------------------------------------------------------------------------------------------------------------------------------------------

%----------------------------------------------- Figure 5: HI+12CO PV diag LONG---------------------------------------------------------
%------------------------------------------------------------------------------------------------------------------------------------------------------

%%%%%%%%%%%%%%%%%%%%% 3.3 Looking for gas flows %%%%%%%%%%%%%%%%%%%%%%%%
\subsection{W43, at the meeting point of large-scale cloud ensembles}
\label{s:coflow}

We further investigated the link between the H$_2$ and \ion{H}{I} gas in the W43 molecular complex and its surroundings by comparing the $^{12}$CO 1--0 CfA and \ion{H}{I} VGPS datacubes. Figures~\ref{fig:co_hi_pvdiag1}-\ref{fig:co_hi_pvdiag2} display the position-velocity diagrams of the W43 molecular complex and its $^{12}$CO/\ion{H}{I} envelope (area outlined by a white dashed rectangle in Fig.~\ref{fig:co_hi_intmap}) plotted against the Galactic latitude and longitude, respectively. According to the definition of \citet{nguyen11b}, we used longitude and latitude extents of $l=29\degr$ to $32\degr$, $b=\pm1\degr$, embracing the $^{12}$CO/\ion{H}{I} envelope of the $^{13}$CO complex. Figures~\ref{fig:co_hi_pvdiag3}a-b complement our investigation of the \ion{H}{I} and $^{12}$CO datacubes for two other latitude ranges. As in Sect.~\ref{s:hienv}, we integrated the $^{12}$CO and \ion{H}{I} lines over the $\vlsr=60-120~\kms$ velocity range.

In Fig.~\ref{fig:co_hi_pvdiag1}a, the $^{12}$CO emission displays two peaks within the white rectangle covering the W43 complex, one at $\sim$95~$\kms$ and another at $\sim$78~$\kms$, recalling the two velocity peaks of Fig.~\ref{fig:co_hi_spec}. In addition, there is a component  of lower density gas, located at more negative latitude (from $b=-1.4\degr$ to $b=-0.4\degr$) and at $\sim$75$\pm10~\kms$. This component remains prominent in the immediate surroundings of W43 and displays a gradient of $\sim$0.2~$\kms \, {\rm pc^{-1}}$ over $\sim$40~pc in the direction of increasing Galactic latitude (see Fig.~\ref{fig:co_hi_pvdiag1}b). This component is associated with a $^{12}$CO velocity gradient of $\sim$0.1~$\kms \, {\rm pc^{-1}}$ over $\sim$400~pc in the direction of increasing longitude (see Fig.~\ref{fig:co_hi_pvdiag3}a and Fig.~\ref{fig:co_hi_pvdiag2}). It is hereafter labelled as cloud ensemble \#1 (see also Figs.~\ref{fig:co_hi_pvdiag1}-\ref{fig:co_hi_pvdiag3}). When approaching the location of W43, the velocity of cloud ensemble \#1 becomes closer to the main velocity range of W43. The $^{12}$CO dispersion in velocity also becomes narrower (see Figs.~\ref{fig:co_hi_pvdiag1}b and \ref{fig:co_hi_pvdiag3}a), suggesting that it could be subject to the increasing gravitational influence of W43. This negative latitude component is associated with enhanced \ion{H}{I} emission (see Fig.~\ref{fig:co_hi_pvdiag1}b) though the velocity gradients, suggested by the $^{12}$CO position-velocity diagrams, are generally not as clearly defined in the \ion{H}{I} position-velocity diagrams. 

Two other $^{12}$CO cloud ensembles, labelled \#2  and  \#3, appear to extend in the direction of increasing Galactic longitude (see Figs.~\ref{fig:co_hi_pvdiag2} and \ref{fig:co_hi_pvdiag3}b). They correspond to the two main peaks of the latitude-velocity diagram integrated over the $l=27\degr-29\degr$ longitude range (see Fig.~\ref{fig:co_hi_pvdiag1}b) and the two line components observed in the $60-120~\kms$ range for both \ion{H}{I} and $^{12}$CO (see Fig.~\ref{fig:co_hi_spec}). The comparison of Figs.~\ref{fig:co_hi_pvdiag3}a and \ref{fig:co_hi_pvdiag3}b shows that they are mostly confined within the latitude range of the W43 molecular complex (from $b=-0.5\degr$ to $b=0.3\degr$) and inspection of Fig.~\ref{fig:co_hi_pvdiag2} shows that they almost disappear in the $l=32\degr-34\degr$ longitude range. The velocity gradients of these $^{12}$CO cloud ensembles are directly linked to the velocity structure of the W43 complex as seen from the $^{12}$CO position-velocity diagram of Fig.~\ref{fig:co_hi_pvdiag3}b. This is even more obvious when inspecting the $^{13}$CO 1--0 longitude-velocity diagram of \citet[][see their Fig.~6]{nguyen11b} and the higher angular resolution $^{13}$CO 2--1 diagram of \citet[][see their Fig.~5a]{carlhoff13}. Both authors have proposed that the velocity component observed at $60-78~\kms$ is remotely associated with the W43 molecular complex, with a possible distance between them of a few hundred parsecs. Cloud ensemble \#2 begins at lower longitude and smaller velocities and  reaches the W43 position and velocity range with an approximate velocity gradient of $\sim$0.02~$\kms \, {\rm pc^{-1}}$ over a distance of $\sim$450~pc  (see Figs.~\ref{fig:co_hi_pvdiag2} and \ref{fig:co_hi_pvdiag3}b). The $^{12}$CO cloud ensemble \#3 is itself found at $\sim$97$\pm7~\kms$, extending for a distance of  900~pc (from $l\sim25\degr$ to $l\sim34\degr$), with several clouds peaking in the $\sim$92-100$~\kms$ range (see Fig.~\ref{fig:co_hi_pvdiag2}). These two $^{12}$CO cloud ensembles are associated with enhancements of the \ion{H}{I} emission which generally are located at $\sim$5~$\kms$ lower velocities than the $^{12}$CO clouds (see Figs.~ \ref{fig:co_hi_pvdiag3}a-b).

%----------------------------------------------- Figure 6: Mass surface density ---------------------------------------------------------
\begin{figure*}[htbp]
\begin{center}
\includegraphics[width=15cm,angle=0]{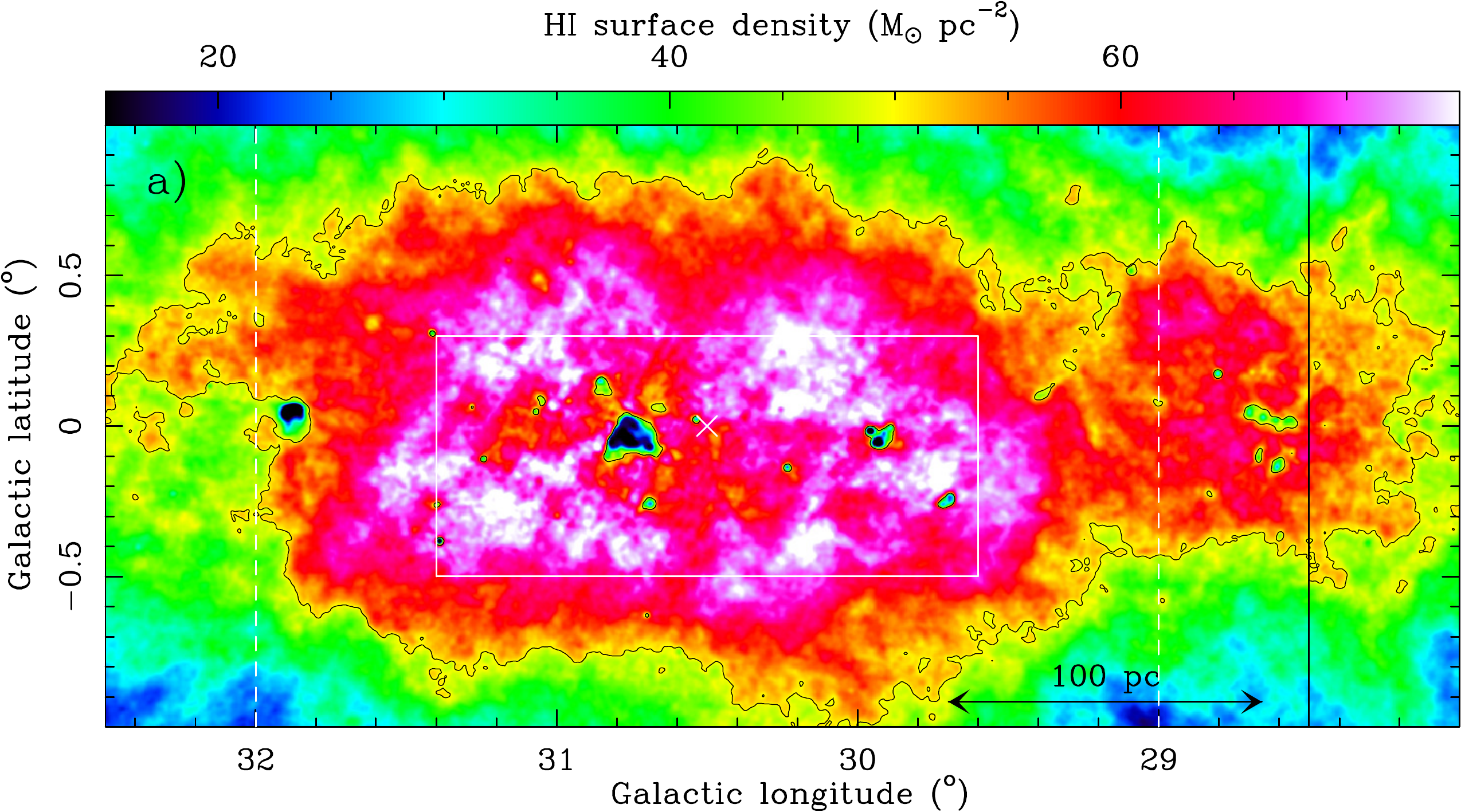}\\
\vskip 0.5cm
\includegraphics[width=15cm,angle=0]{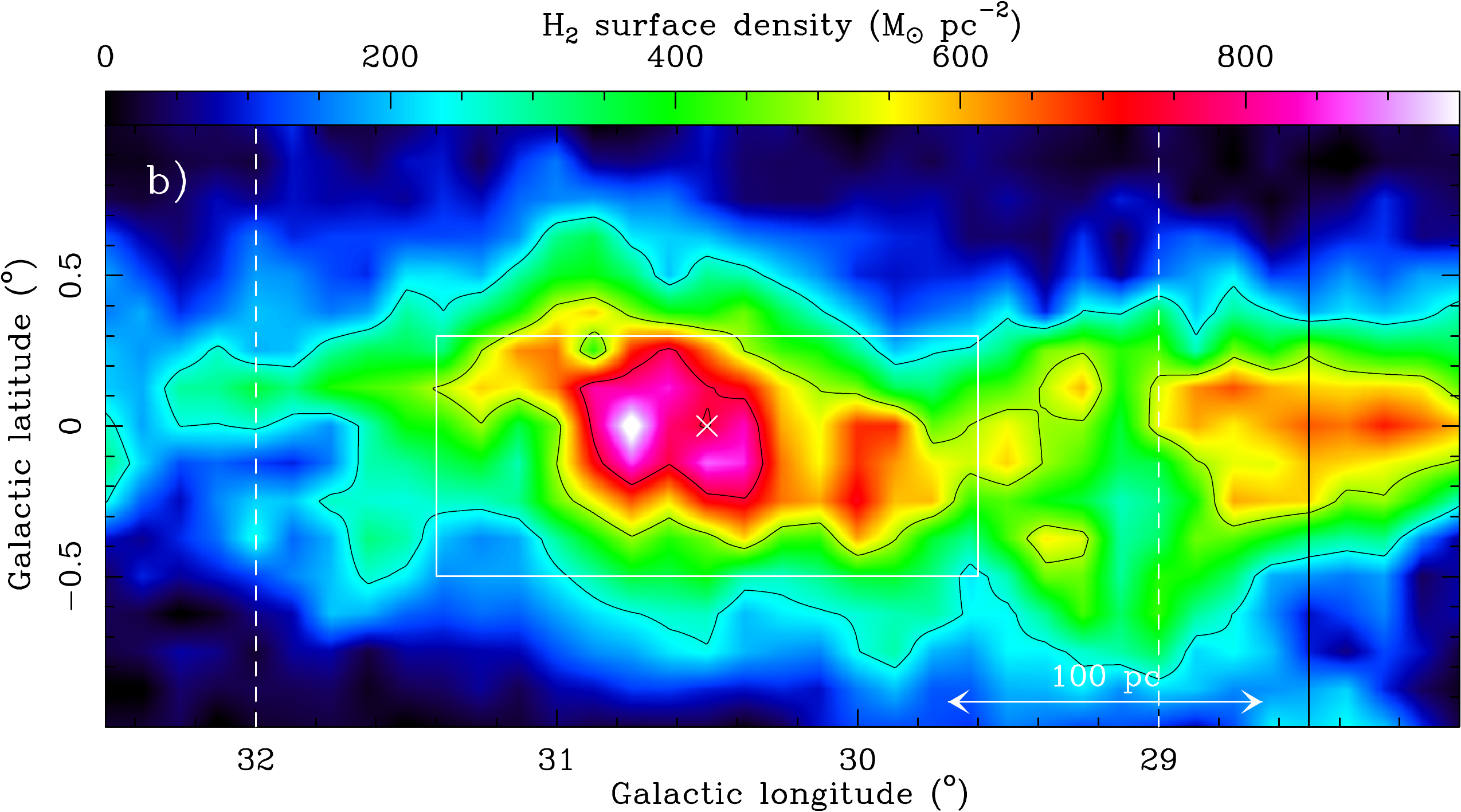}
\end{center}
\vskip -0.3cm
\caption{W43 surface density images of \ion{H}{I} and H$_2$ gases in {\bf a} and {\bf b}, respectively. They were calculated using \ion{H}{I} and $^{12}$CO~1--0 lines integrated over $60-120~\kms$ and Eq.~1 and 3. Contours are $50~\msun\,$pc$^{-2}$ in {\bf a} and 250, 500, and $750~\msun\,$pc$^{-2}$ in {\bf b}. The white boxes outline the extent of the W43 molecular complex (continuous lines) and its \ion{H}{I} envelope (dashed lines) as defined in \citet{nguyen11b}. The black box locates the largest ring used in the  $R_{{\rm H_{2}}}$ diagram (see Fig.~\ref{fig:gasratio1}) and the white cross its center.}
\label{fig:hih2_surfdens}
\end{figure*}
%------------------------------------------------------------------------------------------------------------------------------------------------------

These three velocity gradients, indicated with arrows in Figs.~\ref{fig:co_hi_pvdiag1}-\ref{fig:co_hi_pvdiag3}, are associated with ensembles of clouds which are clear peaks in the position-velocity diagrams. They are not a direct proof of converging/colliding clouds since they could just underline stable clouds following the structure of the Milky Way in the region surrounding W43. Nevertheless, it is tempting to further check if they could be the remnants of \ion{H}{I} gas streams and/or $^{12}$CO streaming clouds that could have collided to form the W43 molecular complex.

%%%%%%%%%%%%%%%%%%%%%%%%%%%%%%%%%%%%%%%%%%%%%%%%%%%%%%%%%
%%%%%%%%%%%%%%%%%%%%% 4. Discussion1 %%%%%%%%%%%%%%%%%%%%%%%%%%%%%
\section{Discussion: Tracing back the formation of the W43 molecular complex}
\label{s:discu}

H$_2$ molecules and \ion{H}{I} atoms are directly linked by formation and destruction processes. In the diffuse parts of cloud complexes, \ion{H}{I} gas forms from the dissociation of H$_2$ molecules by the general interstellar radiation field of the Galactic plane, by the local UV radiation arising from embedded OB clusters, or by  cosmic rays. In contrast, the high-density regions of cloud complexes offer enough shielding from \ion{H}{I} atoms (cold and hot WNM), H$_2$ molecules, and dust so that \ion{H}{I} atoms efficiently turn into H$_2$ molecules on the dust grains surfaces. Observing the formation of H$_2$ gas and molecular clouds from \ion{H}{I} in an envelope or streams is therefore a challenge.

%%%%%%%%%%%%%%%%%%%%% 4.1 HI to H2 transition %%%%%%%%%%%%%%%%%%%%%%%%%%%
\subsection{A surface density threshold for the atomic gas?}
\label{s:hisaturation}

Hypothetically, if there is equilibrium between the formation and destruction of H$_2$, there should be a steep transition from atomic to molecular gas associated with the saturation of the \ion{H}{I} surface density. A value of $\Sigma_{\ion{H}{I}} \sim 10~\msun\,$pc$^{-2}$ has been observed in extragalactic studies \citep[see e.g.][]{BlRo06,bigiel08, wong09}. Similar saturation levels have recently been found for diffuse and dense Galactic clouds \citep[e.g.][]{barriault10,lee12}: $\Sigma_{\ion{H}{I}} \sim 4-7~\msun\,$pc$^{-2}$, equivalent to $N_{\ion{H}{I}} \sim 5-8 \times 10^{20}$~cm$^{-2}$. The equilibrium model of \citet{krumholz09} does predict such a surface density threshold at $\Sigma_{\ion{H}{I}} = 10~\msun\,$pc$^{-2}$ for solar metallicity clouds. In contrast, the dynamical models accrete streams/clouds of \ion{H}{I} and H$_2$ gases to form molecular clouds/complexes which are in neither kinematic nor chemical equilibrium. As a consequence, dynamically-formed molecular clouds should display several transitions from atomic to molecular media along each line-of-sight. The model of e.g.\ \citet[][]{glover10} predicts that the \ion{H}{I} surface density is high when the cloud is assembling, that it then sharply drops within $1-2$~Myr, but that it may never reach equilibrium by $20-30$~Myr. 

To measure the surface density map of atomic gas in W43 (see Fig.~\ref{fig:hih2_surfdens}b), we used the \ion{H}{I} VGPS data integrated over the $60-120~\kms$ range as in Fig.~\ref{fig:co_hi_intmap} and Sect.~\ref{s:coflow}. We assumed that the \ion{H}{I} emission was mainly optically thin and took a distance of 5.5~kpc for W43 and its surroundings. The equation of \citet{spitzer78} for the column density turns into the following equation for the \ion{H}{I} surface density, $\Sigma_{\ion{H}{I}}$, as a function of the \ion{H}{I} line brightness, $\tmb$(\ion{H}{I}):
\begin{equation}
\Sigma_{\ion{H}{I}} \simeq 0.014~\msun\,\mbox{pc}^{-2} \times \frac{\int^{120}_{60} \tmb(\mbox{\ion{H}{I}})~\mbox{d}v} {1\mbox{K\,}\kms}.
\end{equation}

\begin{figure}[htbp]
\includegraphics[width=8.5 cm,angle=0]{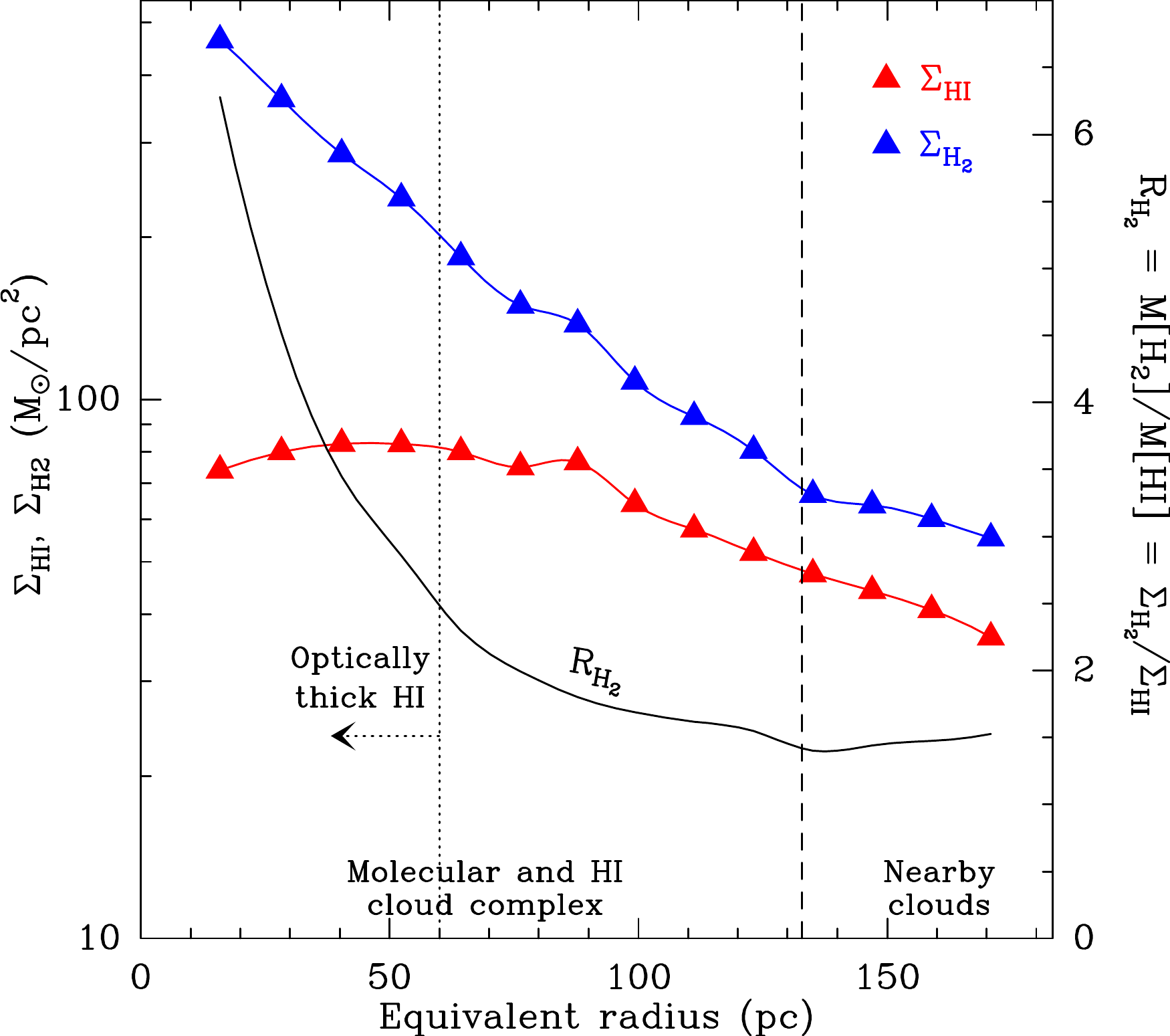}
\caption{
The \ion{H}{I} and H$_2$ surface densities (red and blue curves/triangles respectively) and the H$_2$-to-\ion{H}{I} ratio (black curve) measured throughout the W43 atomic/molecular cloud complex and the nearby clouds. Measurements are averaged within rectangular annuli centered at $l=30.5\degr$ and $b=0\degr$ and with an aspect ratio 3:2. The dashed vertical line outlines the extent of the W43 complex, the dotted vertical line and arrow locate the region inside which the \ion{H}{I} line emission is widely self-absorbed. The \ion{H}{I} surface density is significantly higher than the equilibrium models predict, while the H$_2$  density remains high, in disagreement with a clear \ion{H}{I}/H$_2$ transition.
}
\label{fig:gasratio1}
\end{figure}

Figure~\ref{fig:gasratio1} assumes elliptical geometry (see Sect.~\ref{s:hienv}) and displays the mean \ion{H}{I} surface density across the W43 atomic/molecular complex and its surroundings. We computed $\Sigma_{\ion{H}{I}}$ in rectangular annuli centered at $l=30.5\degr$ and $b=0\degr$, with an aspect ratio 3:2, separated by $\sim$15~pc and $\sim$10~pc along the Galactic longitude and latitude, and reaching up to regions covering the cloud ensembles identified in Sect.~\ref{s:coflow}. In regions where the \ion{H}{I} lines are clearly optically thick, the surface density of \ion{H}{I} is underestimated. Outside these areas, i.e.\ at the $60-170$~pc radii of Fig.~\ref{fig:gasratio1}, $\Sigma_{\ion{H}{I}}$ steadily decreases with radius from $\sim$82~$\msun\,$pc$^{-2}$  to $\sim$36~$\msun\,$pc$^{-2}$. While the gradient of the atomic gas surface density is robust, its absolute value is uncertain by at least a factor of $\sim$2. The main reason is that we have estimated $\Sigma_{\ion{H}{I}}$ by integrating the \ion{H}{I} lines over the $^{13}$CO velocity range which may not perfectly apply for \ion{H}{I} (see Sect.~\ref{s:coline}). Since the line widths of \ion{H}{I} and CO gases in clouds are generally found to be similar \citep[e.g.][]{barriault10}, a factor of $\sim$2 uncertainty seems reasonable. Where the emission is optically thick, this introduces additional uncertainty, especially in the inner parts of W43, at radii smaller than 60~pc. 

%----------------------------------------------- Figure 7: H2-to-HI ratio --------------------------------------------------------------------
%------------------------------------------------------------------------------------------------------------------------------------------------------

The surface density of the \ion{H}{I} gas in W43 is in clear disagreement with previous saturation findings in nearby Galactic clouds \citep[compare the red curve of Fig.~\ref{fig:gasratio1} to e.g.][]{lee12} and resolved galaxies \citep{bigiel08}, as well as with the threshold predicted by \citet{krumholz09}. Indeed, $\Sigma_{\ion{H}{I}}$ is about 6.5 times larger than $10~\msun\,$pc$^{-2}$ and displays a negative radial gradient. The modelled \ion{H}{I} threshold $\Sigma_{\ion{H}{I}} = 10~\msun\,$pc$^{-2}$ is strictly valid for the idealized case of a spherical molecular cloud with solar metallicity and where \ion{H}{I} and H$_2$ gases are mutually exclusive. \citet{krumholz09} extrapolate it to the case of more structured molecular entities, such as molecular complexes composed of several clouds which host OB clusters. They showed that $\Sigma_{\ion{H}{I}}$ is, to a good approximation, independent of the strength of the interstellar radiation field and sublinearly dependent on the metallicity. The W43 molecular complex, with a slightly higher metallicity \citep[possibly $1.4 \times Z_\odot$,][]{nguyen11b} and a very strong mean interstellar radiation field ($20 \times G_0$ according to \citealt{shibai91} and possibly $100-1000 \times G_0$, Schneider, priv.\ com.), is therefore predicted to display an \ion{H}{I} threshold around $10~\msun\,$pc$^{-2}$. Note that, when focusing on a single 10~pc cloud such as W43-Main, the measured \ion{H}{I} surface density decreases by a factor of a few but is still a factor of $\sim$3 times larger  than $10~\msun\,$pc$^{-2}$. The large \ion{H}{I} surface densities observed here are a consequence of either numerous H$_2$ to \ion{H}{I} transitions along each line of sight or a non-steady state formation like that suggested by the \citet{GlML07} models. In the case of W43, both processes probably need to be taken into account. We recall that the W43 molecular cloud complex remains a coherent and gravitationally bound cloud structure, even if it is structured as several individual clouds and numerous layers of H$_2$ and \ion{H}{I} gas.

%%%%%%%%%%%%%%%%%%%%% 4.2 HI to H2 transition %%%%%%%%%%%%%%%%%%%%%%%%%%%
\subsection{The transition from atomic to molecular gas}
\label{s:hitoh2}

The transition from atomic to molecular gas can also be traced by the evolution of the H$_2$-to-\ion{H}{I} ratio from the diffuse \ion{H}{I} envelope to the dense molecular cloud. We estimated the mean H$_2$-to-\ion{H}{I} ratio, $R_{{\rm H_{2}}}$, throughout W43 and its surroundings by computing the mass surface density ratio of molecular H$_2$ and atomic \ion{H}{I} gas. The \ion{H}{I} surface density map shown in Fig.~\ref{fig:hih2_surfdens}a and computed from Eq.~1 has been smoothed to the $450\arcsec$ (or 12~pc) resolution of the CfA data. Associated uncertainties are discussed in Sect.~\ref{s:hisaturation}.

\begin{figure}[htbp]
\includegraphics[width=9 cm]{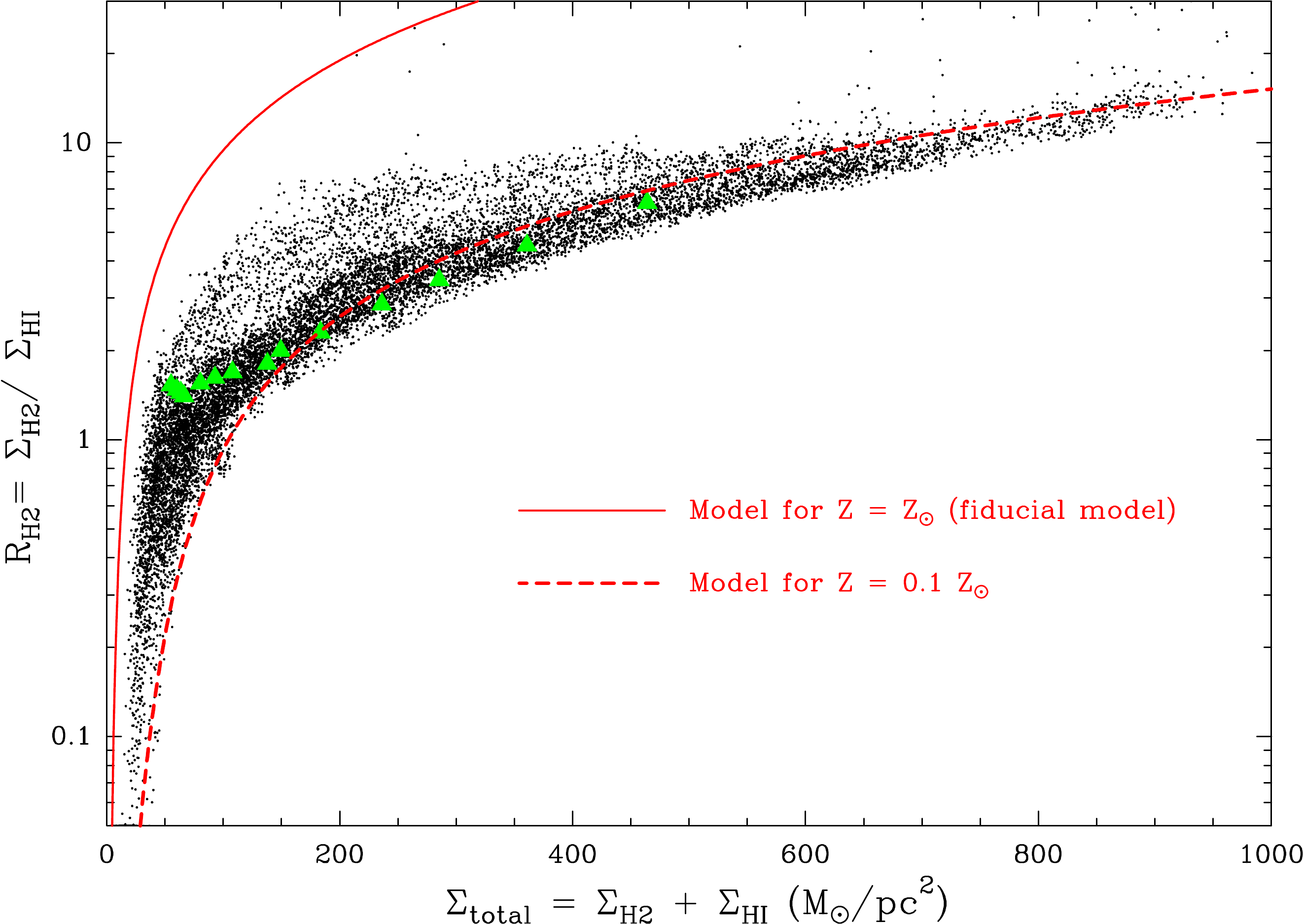}
\caption{The H$_2$-to-\ion{H}{I} ratio, $R_{{\rm H_{2}}}$, of each map point as a function of the total surface density, $\Sigma_{\ion{H}{I}}+\Sigma_{{\rm H_{2}}}$, compared with the predictions of the \citet{krumholz09} model (red curves, taken from their Eq.~39). The green triangles indicate the mean $R_{{\rm H_{2}}}$ values measured within the rings used in Fig.~\ref{fig:gasratio1}. The $R_{{\rm H_{2}}}$ ratio values measured throughout the W43 molecular complex and its surroundings are in striking disagreement with the fiducial model of \citet{krumholz09}.
}
\label{fig:gasratio2}
\end{figure}

We determined the H$_2$ mass surface density from the total intensity of $^{12}$CO 1--0 CfA data, $\tmb(^{12}\mbox{CO})$,  integrated again over $60-120~\kms$. Adopting a value for the CO-to-H$_2$ conversion factor (also known as the X-factor) given by \citet{bloemen86}, the molecular column density, $N_{{\rm H_{2}}}$, is estimated through
\begin{equation}
N_{{\rm H_{2}}} = 2.75 \times 10^{20}\,\mbox{cm}^{-2}\, 
\times\, \frac{\int^{120}_{60} \tmb(^{12}\mbox{CO})~\mbox{d}v} {1\mbox{K\,}\kms}~.
\end{equation}
With a mean molecular weight of $\mu= 2$, the mass surface density of molecular gas, $\Sigma_{{\rm H_{2}}}$, can be computed by the following equation: 
\begin{equation}
\Sigma_{{\rm H_{2}}} \simeq  4.41~\msun\,\mbox{pc}^{-2} \times \frac{ \int^{120}_{60} \tmb(^{12}\mbox{CO})~\mbox{d}v }{1\mbox{K\,}\kms}.
\end{equation}
Figure~\ref{fig:hih2_surfdens}b gives the surface density map of molecular gas traced by $^{12}$CO 1--0. The H$_2$ surface density values are uncertain by at least a factor of two because the X-factor, converting CO emission into H$_2$ mass, is expected to vary across the Milky Way. According to \citet{shetty11}, the X-factor only makes sense as a cloud-scale average. However, the relations observed in Figs.~\ref{fig:gasratio1}-\ref{fig:gasratio2} should be secure since cloud properties are investigated through $^{12}$CO emission averaged over one to one hundred beams of HPBW~$\simeq 13$~pc. Moreover, given that W43 lies in the inner part of the Galactic disk, the amount of CO-dark H$_2$ gas should be minimal. Note that instead of the surface density ratios used by e.g.\ \citet{lee12}, we could have computed mass ratios, since they are strictly equal:
\begin{equation}
R_{{\rm H_{2}}} =  f_{{\rm H_{2}}} / f_{\ion{H}{I}} = M_{{\rm H_{2}}}/M_{\ion{H}{I}} = \Sigma_{{\rm H_{2}}}/\Sigma_{\ion{H}{I}},
\end{equation}
where the H$_2$ and \ion{H}{I} gas fractions are defined by $ f_{{\rm H_{2}}} = M_{{\rm H_{2}}}/(M_{\ion{H}{I}}+M_{{\rm H_{2}}})$ and $ f_{\ion{H}{I}} = M_{\ion{H}{I}}/(M_{\ion{H}{I}}+M_{{\rm H_{2}}})$. 

In Fig.~\ref{fig:gasratio1}, we see that the H$_2$-to-\ion{H}{I} ratio smoothly rises with decreasing radii with values of $R_{{\rm H_{2}}} \sim 1.5$ at the outskirts of the complex (i.e. at a radius of 135~pc) and $\sim$6 at 16~pc from the center of the complex. Even if we exclude the regions inside which the \ion{H}{I} lines are clearly optically thick, and therefore focusing on the outer parts of what we called the \ion{H}{I} envelope in Sect.~\ref{s:hienv}, we see that the $R_{{\rm H_{2}}}$ ratio still increases by about a factor of two as we move toward the center of the complex. Figure~\ref{fig:gasratio1} therefore suggests that \ion{H}{I} turns into  H$_2$ more efficiently at greater depths within the W43 complex. The surroundings of the W43 complex (called nearby clouds in Fig.~\ref{fig:gasratio1}) display a slight increase of $R_{{\rm H_{2}}}$ from the $\sim$1.5 value measured at the outskirts of the complex. This behavior marks the start of new molecular cloud structures, confirming our definition of the W43 atomic/molecular complex.

The transition from \ion{H}{I}-dominated to H$_2$-dominated gas is usually set at either $R_{{\rm H_{2}}}=0.25$ or $R_{{\rm H_{2}}}=0.1$ \citep[see e.g.][and references therein]{lee12}. With either definition, the observed molecular ratio (black curve in Fig.~\ref{fig:gasratio1}) surprisingly suggests that this transition happens outside the \ion{H}{I} envelope that we have identified for W43. As previously mentioned, the absolute values of $\Sigma_{\ion{H}{I}}$ (or $M_{\ion{H}{I}}$) and $\Sigma_{{\rm H_{2}}}$ (or $M_{{\rm H_{2}}}$) are both uncertain by factors of at least two. Even when accounting for these uncertainties (i.e.\ corresponding to values of $R_{{\rm H_{2}}}$ correct to within a factor 4), we cannot reconcile the molecular ratio of Fig.~\ref{fig:gasratio1} and our picture of a molecular complex surrounded by an \ion{H}{I} envelope (see Sect.~\ref{s:hienv}). Extending Fig.~\ref{fig:gasratio1} to larger radii would not make sense since wider annuli would cover gas from other molecular structures observed at greater distances (see Fig.~\ref{fig:co_hi_intmap} and Sect.~\ref{s:nearbyclouds}). 
 
%----------------------------------------------- Figure 8: HI-H2 transition --------------------------------------------------------------------
%------------------------------------------------------------------------------------------------------------------------------------------------------

If we take the $\Sigma_{\ion{H}{I}}$ and $R_{{\rm H_{2}}}$ curves of Fig.~\ref{fig:gasratio1} at face value, the W43 complex seems to develop/form in an area globally shielded against the Galactic radiation field. This is in direct agreement with the high global molecular mass fraction of W43 \citep[$\sim$65\%,]{nguyen11b} which follows the general increase with the Galactic radius, observed for other molecular clouds (Elmegreen \& Elmegreen 1987). Figure~\ref{fig:gasratio2} plots the value of $R_{{\rm H_{2}}}$ for each map point within the rectangle outlined by the black lines in Fig.~\ref{fig:co_hi_intmap} and therefore containing both the W43 complex and nearby clouds. As a function of their total surface density ($\Sigma_{\ion{H}{I}}+\Sigma_{{\rm H_{2}}}$), the distribution of points cannot be fit by the fiducial model of \citet{krumholz09} with a solar metallicity, and instead requires a much smaller and unrealistic metallicity: $0.1 \times Z_\odot$ rather than the value of $1.4 \times Z_\odot$ found by \citet{nguyen11b}. It also confirms that the W43 region, including the \ion{H}{I} envelope, globally lies in a domain already shielded from external sources of UV radiation. Indeed, Fig.~\ref{fig:gasratio2} does not trace the turnover between atomic-dominated and molecular-dominated zones observed by \citet[][see their Fig.~12]{lee12} and predicted by \citet{krumholz09}. The purely-atomic envelope of W43 can be seen at high latitude but seems not to exist within the Galactic disk. This can be explained by the high gas and stellar pressure that should exist in the W43 region. Indeed, \cite{BlRo06} showed that, in galaxies, there is a relation between the combined pressures from stars and gas and the molecular ratio of clouds. Gas structures associated with the velocity gradients found in Sect.~\ref{s:coflow} are molecular and have a complex structure well before any interaction with W43, which argues that they should  be called cloud ensembles rather than gas streams.

From Fig.~\ref{fig:gasratio1}, we also see that $\Sigma_{\ion{H}{I}}$ and $R_{{\rm H_{2}}}$ simultaneously increase within the outer part of the \ion{H}{I} envelope as the radius decreases from 135~pc to 60~pc. This increase of the \ion{H}{I} surface density suggests a continuous accumulation of  \ion{H}{I} gas in this heterogeneous H$_2$/\ion{H}{I} structure, even if it is globally shielded from the Galactic field and protected from the UV field of the embedded \ion{H}{II} regions. In this scenario the \ion{H}{I} and H$_2$ gases are well-mixed, coexisting but far from chemical equilibrium, as proposed by the models of e.g.\  \citet{HeHa08} or \citet{clark12}. The fact that some H$_2$ is likely being converted back into \ion{H}{I} is demonstrated by the strong ($3\times 10^5~\lsun$) and large-scale ($1\degr\times1\degr$) C~{\scriptsize II} emission extended over the complete W43 complex, which is most probably associated with its many photodissociation layers \citep{shibai91}. The increase in the H$_2$-to-\ion{H}{I} ratio itself traces the global formation of molecular clouds thanks to H$_2$ molecules forming and the molecular medium becoming denser. This also happens far from the area already defined as the W43 molecular cloud complex. If confirmed, these two results point toward a dynamical and non-steady state scenario for the formation of the W43 molecular complex.

%%%%%%%%%%%%%%%%%%%%% 4.3 Nearby clouds %%%%%%%%%%%%%%%%%%%%%%%%%%%
\subsection{Clouds following the Galactic structure/kinematics}
\label{s:nearbyclouds}

We have identified, in the immediate surroundings of the W43 molecular complex, three cloud ensembles associated with structured $^{12}$CO and \ion{H}{I} emission (see Fig.~\ref{fig:co_hi_pvdiag2} and Sect.~\ref{s:coflow}). With a common distance of 5.5~kpc, these three ensemble of clouds are developing $\sim$300~pc away from W43, while their spatial and velocity distributions suggest that they join at the location of  the W43 molecular complex (see arrows in Figs.~\ref{fig:co_hi_pvdiag1}-\ref{fig:co_hi_pvdiag3}). To position these clouds in the W43 region, one should remember that W43 is located close to the tangent point of the Scutum-Centaurus arm and probably at/close to its connecting point with the long bar of the Milky Way.

Cloud ensemble \#1 originates from longitudes smaller than that of W43 and approaches W43 from below the Galactic plane (see Figs.~\ref{fig:co_hi_pvdiag1}b, \ref{fig:co_hi_pvdiag2}, and \ref{fig:co_hi_pvdiag3}a). It could well be a high-latitude cloud attracted by the gravitational potential well of the Galactic plane and W43.

The two main cloud ensembles (\#2 and \#3) are themselves confined at low latitude and are seen at longitudes smaller than that of W43: from $b=-0.5\degr$ to $b=0.3\degr$ and from $l=27\degr$ to $l=30\degr$ (see Figs.~\ref{fig:co_hi_pvdiag1}-\ref{fig:co_hi_pvdiag2}). Cloud ensemble \#2 follows a velocity gradient close to that found for the leading part of the Scutum-Centaurus arm, shown schematically by a thick gray curve in Fig.~\ref{fig:co_hi_pvdiag2} \citep[e.g.][]{vallee08,DT11}. Cloud ensemble \#3 has a velocity gradient consistent with that of the preceding part of the Scutum-Centaurus arm as defined by \citet[][]{vallee08} (see Fig.~\ref{fig:co_hi_pvdiag2}). However, the new model of the Milky Way by e.g.\  \citet{DT11} stops the spiral structure of the Scutum-Centaurus arm at the location of W43 and starts the long bar a few degrees away \citep{lopez07,benjamin05,RFCo08}. Therefore, cloud ensemble \#3 may alternatively correspond to clouds at the tip of the long bar \citep[see also][]{gonz12}. As shown in Fig.~\ref{fig:co_hi_pvdiag2}, the velocity gradients of cloud ensembles \#2 and \#3 do not lie on the median location of the Scutum-Centaurus arm modeled with circular orbits (compare the arrows and the thick gray curve). We indeed measured $\sim$1$\degr$ and up to $\sim$$20~\kms$ offsets between the observed cloud ensembles and the position-velocity structure predicted by \citet{vallee08}. Such offsets for a single cloud could easily be explained by streaming motions of gas in and out of the gravitational well of spiral arms. However, in the case of clouds spanning $\sim$500~pc along the Galactic plane through and around W43, we favor an explanation related to the interaction with the long bar. Cloud ensembles \#2 and \#3 could therefore simply follow the structure of the Milky Way without interacting much with each other or with W43. Nevertheless, their location near the tip of the Galactic bar makes them subject to both the spiral arm circular rotation and the solid-body bar elliptical rotation. According to numerical simulations, such locations in galaxies have a high probability of cloud-cloud collision due to orbit crowding \citep[e.g.][]{RFCo08}.

%%%%%%%%%%%%%%%%%%%%% 4.4 Colliding clouds %%%%%%%%%%%%%%%%%%%%%%%%%%%
\subsection{Clouds streaming toward W43?}
\label{s:collidingclouds}

With the refined 5.5~kpc distance and the velocity range discussed in Sect.~\ref{s:coline}, we have recomputed the mass of the W43 complex, whose Galactic limits have been set by \cite{nguyen11b} (see white rectangle of 130~pc equivalent diameter in Fig.~\ref{fig:co_hi_intmap}). Without any correction for optical thickness, the atomic \ion{H}{I} and molecular H$_2$ $+$ He gas masses of W43 are $M_{\mbox{\rm{\tiny \ion{H}{I}}}} \sim 9\times 10^5~\msun$ and $M_{\rm H_2 + He} \sim 8\times 10^6~\msun$, leading to a total gas mass of $M_{\mbox{\rm \tiny W43}} \sim 9\times 10^6~\msun$. With the same assumptions, the mass of the \ion{H}{I} envelope surrounding W43 (see white dashed rectangle of 270~pc equivalent diameter in Fig.~\ref{fig:co_hi_intmap}) is $M_{\mbox{\rm \tiny \ion{H}{I}, env}} \sim 3\times 10^6~\msun$. With $\sim$$10^7~\msun$ of gas, the W43 $^{12}$CO/\ion{H}{I} complex is a large gravitational potential well that may attract nearby molecular clouds of $\sim$$10^5~\msun$ mass. Here we briefly investigate the kinematic state of the three cloud ensembles  identified in Sect.~\ref{s:coflow}, with the aim of evaluating whether they are streaming toward W43.

The fact that these three cloud ensembles meet at the location of W43 could reflect a real physical interaction. A first argument is that the velocity dispersion measured for W43 is large (FWHM$_{^{13}\rm CO}\sim 22.3~\kms$; see \citealt{nguyen11b}) and could better account for agglomeration of clouds than micro-turbulence. Such a phenomenon is called gravitationally driven turbulence in dynamical models of cloud formation \citep[see e.g.][]{heitsch08}. Cloud ensembles \#1 to \#3 could thus be streams that are accreted onto the central W43 region, driven simply by gravity. Such accretion from the ambient surrounding gas, at a late stage in the formation of molecular cloud complexes, has been noted in the numerical models of e.g.\ \citet{vazquez07},  \citet{HeHa08}, and \citet{dobbs08}.

%Rotation vs. infall
Velocity gradients of molecular clouds are often interpreted as tracing their specific angular momentum, which can be compared with Galactic rotation models and predictions of cloud formation scenarios \citep[e.g.][]{ImBl11}. We do not investigate this idea here since cloud ensembles \#1 to \#3 are defined on much larger scales than that of individual clouds and  ensembles \#2 and \#3 tend to reflect the velocity structure of the Milky Way (see Sect.~\ref{s:nearbyclouds}). This last point implies that the major part of the gradients observed in the direction of increasing longitude does not trace the infall velocities of the cloud ensembles onto W43. Since cloud ensemble \#1 is mainly observed in the direction of decreasing latitude, its velocity gradient could more meaningfully be compared to its predicted free-fall velocity toward the W43 $^{12}$CO complex. In fact, the $\sim$0.2~$\kms\, {\rm pc^{-1}}$ gradient over $\sim$40~pc of Galactic latitude is consistent with the free-fall velocity of clouds located $\sim$200~pc away from the $\sim$$10^7~\msun$ W43 complex. Interestingly, the velocity dispersion of cloud ensemble \#1, estimated through $^{12}$CO line widths, narrows when approaching W43, with the FWHM decreasing from $\sim$$20~\kms$ to $<$$10~\kms$ (see Fig.~\ref{fig:co_hi_pvdiag3}a), suggesting again that it is attracted by the deep gravitational potential well of W43. We do not see any acceleration of the infall with decreasing distance of the cloud ensemble \#1 to W43 but this can easily be covered up by projection and rotation effects as well as cloud inhomogeneities. As for cloud ensembles \#2 and \#3, they are initially separated by $>$$40~\kms$ velocities along the line of sight and extend within the W43 complex as velocity structures. The complex structure in velocity of W43 is also testified by its velocity dispersion which is much larger than that of individual cloud ensembles streaming toward it: FWHM$_{^{12}\rm CO}^{\rm W43}\sim 30~\kms$ for W43 vs. $\sim$$10~\kms$ for nearby clouds (see Fig.~\ref{fig:co_hi_pvdiag2}). The $\sim$$20~\kms$ velocity jumps observed in the densest parts of the W43 complex such as the W43-Main and W43-South clouds (see Fig.~\ref{fig:co_hi_pvdiag2}) could represent local and abrupt accelerations of the infall. 

%Shears
According to \cite{carlhoff13}, the W43 $^{13}$CO molecular complex is stable against Galactic shear. The $^{12}$CO/\ion{H}{I} cloud complex, which extends to much greater distances (from the white to the white-dashed rectangle in Fig.~\ref{fig:co_hi_intmap}) and the 100~pc ensemble of clouds located at its outskirts (from the white-dashed to the black rectangle in Fig.~\ref{fig:co_hi_intmap}) should be far more sensitive to Galactic shear. Indeed, despite their huge total mass, up to $\sim$$10^7~\msun$, the large area they cover, $\sim$$6-8\times10^4$~pc$^2$, and their velocity range up to $60~\kms$ lead to a shear parameter of $S_{g}=1-3$ \citep[with Eq.~8 of][]{dib12}. In a simple spiral rotation model of the Milky Way, both the envelope and nearby clouds would thus not be linked to W43. However, numerical simulations show that areas in front of galactic bars are locations where clouds sporadically agglomerate \citep[e.g.][]{athanassoula92,wozniak07, renaud13}. Other indirect arguments come from  extragalactic observational studies that found extreme molecular cloud complexes at the ends of galactic bars \citep[e.g. M83, NGC~1300; see][and references therein]{MaFr97}.

To conclude, we do not find definitive proof that nearby clouds are streaming toward W43, though, as discussed here, numerous factors favor  this interpretation. W43, at a special location in the Milky Way, is denser and more dynamic than typical molecular clouds of  the Galactic disk. These unusual characteristics have allowed us to identify clouds possibly streaming toward W43. Deeper observations and careful analyses of other cloud complexes may teach us whether cloud agglomeration occurs as frequently as it appears to within numerical simulations \citep[see e.g.][]{dobbs08}.

\subsection{History of cloud and star formation in the region of W43}
\label{s:historySF}

% Molecular nature and collision
Here we try to extrapolate the interpretation that cloud ensembles are streaming toward W43  back to times when the W43 molecular cloud complex was just starting to form. According to numerical models, the special location of W43 in the Milky Way should have promoted its ability to accumulate \ion{H}{I} and H$_2$ gas from different structures \citep[e.g.][]{RFCo08}. We have shown, in Sect.~\ref{s:hitoh2}, that most of the H$_2$ molecules form far out of the classical $\sim$10--50~pc outskirts of molecular clouds \citep[e.g.][]{lee12} and further even than the 60--135~pc radii (from the center of the complex) taken to be the optically thin part of the \ion{H}{I} envelope (see Fig.~\ref{fig:gasratio1}). The W43 molecular cloud complex may therefore have formed through the agglomeration of clouds containing both molecular and atomic gas. The clouds that currently are potentially streaming toward W43 (see Sect.~\ref{s:collidingclouds}) display velocity gradients following the Galactic structure. They thus can reasonably be taken as models for the initial cloud flows that created W43 but their surface density could be orders of magnitude higher or lower.

% History
The level of cloud structure and the star formation activity of W43 and its nearby clouds can help define the chronology of the  cloud and star formation events in the region. The molecular cloud structure \#2 qualifies as `pre-star forming' according to \citet{sawa12}, who labeled it cloud M-2. They argue that it  is not yet structured at high density, with a moderate brightness distribution index, and that it hosts very few \ion{H}{II} regions \citep{AB09}. For comparison, W43 (their cloud H-2) is categorized as an `active star-forming complex' since it has a high brightness distribution index and contains many more \ion{H}{II} regions.

Molecular cloud ensemble \#3 was not discussed by \citet{sawa12}. It hosts many \ion{H}{II} regions \citep{AB09} and an enhanced number of Red Super Giant (RSG) clusters/associations \citep[see e.g. Fig.~9 of][]{gonz12}. The three main RSG associations of the first quadrant are indeed located around G29.2-0.2 \citep[RSGC3 $+$ Alicante7 $+$ Alicante10, e.g.][]{GN12}, G26.2-0.1 \citep[RSGC2, also called Ste2, see e.g.][]{negu12}, and G24.6+0.4 \citep[RSGC1 $+$ Alicante8,][]{negu10}. They contain numerous ($30-50$) RSG stars, are thus massive, $\sim$2$-10\times 10^4~\msun$, stellar associations. \citet{gonz12} locate them at the tip of the long bar of the Milky Way and $\sim$6~kpc from the Sun. Their presence at $l=24\degr-30\degr$ suggests that, $\sim$20~Myrs ago, there were several sites of extremely active cloud and star formation located $\sim$500~pc downstream of W43 (see Fig.~\ref{fig:co_hi_intmap}). This spatial offset is reminiscent of those found between clouds and star clusters in e.g. the NGC~6951 galaxy \citep{vdlaan11}. Molecular cloud ensemble \#3 is associated with these RSGCs and is one order of magnitude less massive than W43 (see Fig.~\ref{fig:co_hi_intmap}). We thus propose that cloud agglomeration and star formation have started in the cloud complex that was associated with velocity gradient \#3. It is now enhanced in W43 mainly owing to clouds infalling from cloud ensemble \#2. This chronology is in agreement with numerical simulations of galaxy evolution with a barred potential, which tend to accumulate clouds in spiral leading arms ahead of the bar, form superstar clusters within these cloud complexes and then drag these clusters and clouds within the bar, where they often disrupt \citep[][in prep.]{wozniak07, renaud13}.

%%%%%%%%%%%%%%%%%%%%%%%%%%%%%%%%%%%%%%%%%%%%%%%%%%%%%%%%%
%%%%%%%%%%%%%%%%%%%%% 5. Conclusion %%%%%%%%%%%%%%%%%%%%%%%%%%%%%
\section{Summary and conclusions}
\label{s:conc}

In order to improve our knowledge of molecular cloud formation, we have initiated an investigation into W43, an extreme molecular cloud and star-forming complex located at a dynamical place within the Milky Way, 5.5~kpc from the Sun. We used an \ion{H}{I} and $^{12}$CO 1--0 database allowing us to trace back the formation of molecular clouds on 10~pc to 400~pc scales for this region. Our main findings can be summarized as follows:
 \begin{enumerate}
 \item We detected an \ion{H}{I} envelope around the W43 molecular cloud (see Fig.~\ref{fig:co_hi_intmap}), confirming the work of \citet{nguyen11b}. It has an aspect ratio of 3:2 (with the larger axis along the Galactic longitude), an equivalent diameter of $\sim$270~pc, and a mass of $\sim$$3\times 10^6~\msun$. Interestingly, such a well-defined and symmetrical envelope is rarely seen around star-forming regions.
\item Three cloud ensembles of \ion{H}{I} and $^{12}$CO gas develop along hundreds of parsecs, at the outskirts of and within the atomic envelope (see Figs.~\ref{fig:co_hi_pvdiag1}-\ref{fig:co_hi_pvdiag3}). The velocity gradients of these cloud ensembles are used to position them in the W43 region. Cloud ensemble \#1 may fall from the  high-latitude part of the Galactic disk, \#2 seems to originate from the Scutum-Centaurus arm, and \#3 may be at the very tip of the long bar. This picture is consistent with W43 being in front of the long bar, a region with crowded orbits known to be efficient in accumulating gas.
\item These three (mostly) molecular ensemble of clouds are separated by $\sim$20--40~$\kms$ along the line of sight and converge on the location of W43. The measured velocity gradients extend within the W43 complex as velocity structures, are consistent with free-fall velocities, and display $\sim$20~$\kms$ velocity jumps where they merge. The cloud ensembles \#1--\#3 could thus be streams that are accreted onto the central W43 region, which has a total mass of $\sim$10$^7~\msun$, driven by gravity rather than forced by converging flows of atomic gas.
\item The \ion{H}{I} surface density measured throughout the W43 complex is high when compared to the values usually quoted in the literature:  $\Sigma_{\ion{H}{I}} \sim 65\pm20~\msun\,$pc$^{-2}$ instead of $7-15~\msun$\,pc$^{-2}$. The \ion{H}{I} surface density does not show a saturated behavior but rather increases when entering the molecular complex (see Fig.~\ref{fig:gasratio1}). This  suggests that W43 is a heterogeneous structure of mixed H$_2$ and \ion{H}{I} gas and that equilibrium between H$_2$ formation and photodissociation is not reached. These results thus argue in favor of a non steady-state formation of molecular clouds in W43. 
\item The H$_2$-to-\ion{H}{I} ratio measured over the complete W43 region is high, $R_{{\rm H_{2}}}\sim3.5\pm_{2}^{3}$, proving that most of the gas is already in molecular form before reaching W43 (see Fig.~\ref{fig:gasratio2}). High-density regions of galaxies, such as the W43 molecular cloud complex, located in what was previously called the molecular ring of the Milky Way, obviously cannot provide constraints on the transition from atomic to molecular material. The formation of such molecular cloud entities is therefore weakly related to the efficiency of turning \ion{H}{I} atoms into H$_2$ molecules.
\item We conclude that the W43 molecular cloud complex most probably formed through a dynamical process such as the collision of several clouds, already in a molecular-dominated regime. The geometry and strength of this collision, as well as the molecular ratio and structure of the clouds, should be used to customize colliding cloud models. This is one of the goals of the W43-HERO IRAM large program which uses the IRAM 30~m telescope to determine the diagnostics of colliding clouds and explain the extreme characteristics of the W43 molecular cloud star-forming region.
\end{enumerate}

\begin{acknowledgements}
Part of this work was supported by the French National Agency for Research (ANR) projects `PROBeS' and `STARFICH', numbers ANR-08-BLAN-0241 and ANR-11-BS56-010. This project is carried out within the Collaborative Research Council 956, sub-project A4, funded by the Deutsche Forschungsgemeinschaft (DFG). 
SG acknowledges support from the DFG via SFB project 881 `The Milky Way System' (sub-projects B1, B2 and B8). This publication made use of the synergy between observers and modellers set up for the large IRAM program W43-HERO. We are grateful to Snezana Stanimirovic and Simon Bihr for useful discussions.
\end{acknowledgements}

%------------------------------------------------------------------------------------

\bibliographystyle{aa}
\bibliography{w43_structure.bib}

\end{document}